\documentclass{article}
\usepackage[utf8]{inputenc}

\usepackage{mystyle}
\usepackage{authblk}
\usepackage{subfig}

\title{Hadron-ion collisions in \Pythia and the vector-meson dominance model for photoproduction}
\author[1,2]{Ilkka Helenius}
\author[1,2]{Marius Utheim}
\affil[1]{University of Jyvaskyla, Department of Physics, P.O. Box 35, FI-40014 University of Jyvaskyla, Finland}
\affil[2]{Helsinki Institute of Physics, P.O. Box 64, FI-00014 University of Helsinki, Finland}
\date{}

\newcommand{\ncoll}{n_\mathrm{coll}}
\newcommand{\sigtot}{\sigma_\mathrm{tot}}
\newcommand{\sigabs}{\sigma_\mathrm{abs}}
\newcommand{\sigdiff}{\sigma_\mathrm{diff}}
\newcommand{\sigSD}{\sigma_\mathrm{SD}}
\newcommand{\sigDD}{\sigma_\mathrm{DD}}
\newcommand{\sigND}{\sigma_\mathrm{ND}}
\newcommand{\sigel}{\sigma_\mathrm{el}}
\newcommand{\Pom}{\mathbb{P}}

\renewcommand{\p}{\mathrm{p}}
\renewcommand{\Pb}{\mathrm{Pb}}
\newcommand{\eCM}{E_\mathrm{CM}}

\newcommand{\pTRef}{p_{\mathrm{T},0}^\mathrm{ref}}

\begin{document}

\maketitle

\begin{abstract}
We present an extension to the \Pythia Monte Carlo event generator that enables simulations of collisions between a generic hadron beam on a nuclear target with energy variation in event-by-event basis. This builds upon \Pythia's module for heavy ions, \Angantyr, as well as previous work on simulating hadron-proton collisions. As such, the extensions in this work are largely technical, except for a rudimentary model for hadronic fluctuations. With hadron-ion simulations, we implement an explicit vector-meson dominance (VMD) model that can be used to simulate interactions of hadronic component of real photons in photo-nuclear collisions. Such processes can be studied in ultra-peripheral heavy-ion collisions and in the future also with the upcoming Electron-Ion Collider. Our work also has applications to hadronic showers, e.g. air showers initiated by high-energy cosmic rays. We first validate the VMD model by comparing to HERA photoproduction data on proton target. Then we apply this to generate events for ultra-peripheral heavy-ion collisions at the LHC and present the results corresponding to the event-selection criteria matching to a recent ATLAS analysis. We find that single-particle multiplicity and rapidity distributions are well in line with the measured ones. We also construct the Fourier coefficients from two-particle correlations for the simulated events and study whether the resulting azimuthal anisotropies are consistent with the ATLAS results. 
\end{abstract}

\section{Introduction} 

\Pythia is a general-purpose event generator that can simulate high-energy collisions events \cite{Bierlich:2022pfr}. The underlying physics modelling is built upon collinear factorization \cite{Collins:1989gx} where hard-process cross sections can be factorized from long-distance physics describing the structure of the colliding particles. The latter can be obtained by applying DGLAP evolution for a non-perturbative input that is fitted to experimental data in a global analysis of parton distribution functions (PDFs) \cite{Kovarik:2019xvh}. In event generators the DGLAP equations can be applied also to generate parton showers where highly-virtual partons formed in a hard process radiate more partons  \cite{Hoche:2014rga}, which later on turn into hadrons at non-perturbative scales that construct the complete final state. Within the previous decade the main focus in \Pythia has been in proton-proton collisions performed at the Large Hadron Collider (LHC) at CERN. More recently there has, however, been several developments to extend the modelling also to other beam configurations including photoproduction in lepton-proton collisions \cite{Helenius:2017aqz}, collisions involving heavy ions \cite{Bierlich:2018xfw}, and collisions with varying hadron beam and a proton target \cite{Sjostrand:2021dal}. In this work we combine the two latter features to enable event generation for collisions of a generic hadron and a heavy ion. Such modelling has applications to photon-induced processes such as ultra-peripheral heavy-ion collisions at the LHC and electron-ion collisions in the future Electron-Ion Collider (EIC) \cite{Accardi:2012qut, AbdulKhalek:2021gbh}, as well as hadronic cascades such as air showers initiated by high-energy cosmic rays.

In ultra-peripheral collisions (UPCs) \cite{Bertulani:2005ru, Klein:2020fmr} the beam particles encounter with a large impact parameter such that no short-distance strong interaction, described with quantum chromodynamics (QCD), can take place but long-distance electromagnetic interactions can still occur. These interactions can be described by exchanging a quasi-real photon and thus allow us to study photon-photon and photon-hadron collisions. Unlike in case of charged point-like leptons, the photons emitted by charged hadrons typically have a limited virtuality which essentially follows from the requirement of hadron staying intact when emitting a photon. Therefore these processes are similar to photoproduction \cite{Klasen:2002xb}, studied extensively e.g. in electron-proton collisions at HERA \cite{Butterworth:2005aq}. With heavy-ion collisions at the LHC one can also study interactions between a real photon and a nuclear target at collider energies. Several recent experimental studies have been performed, including exclusive particle production \cite{ALICE:2013wjo, ALICE:2021gpt, CMS:2016itn, LHCb:2021bfl}, inclusive dijets \cite{ATLAS:2022cbd} and two-particle correlations \cite{ATLAS:2021jhn}. Interestingly, the latter study have found signs of long-range correlations typically associated with hydrodynamical flow of strongly interacting medium \cite{Heinz:2013th}. Similar effects have been observed earlier in case high-multiplicity proton-nucleus \cite{CMS:2015yux, ATLAS:2016yzd} and proton-proton \cite{ATLAS:2015hzw, ATLAS:2019wzn} collisions which have raised a question whether there are other mechanisms that could result as similar signatures for collective behaviour. The high-multiplicity photo-nuclear interactions offer thus a complementary ``small'' collisions system to study onset of hydrodynamic behaviour and the origin of the observed collective effects.

When studying collisions of a real photon there are several contributions that should be accounted for. The photons can of course interact as point-like particle (direct photon) or it can fluctuate into a hadronic state (resolved photon) with equal quantum numbers. The latter may happen non-perturbatively, or the photon can split into quark-antiquark pair perturbatively. The partonic structure of these hadronic states can be encoded in PDFs which now contain a perturbatively calculable point-like component and a hadron-like component. As the latter contribution is not known, a non-perturbative input that can be fixed in a global QCD analysis is needed. Another approach for this is to apply the so called vector-meson dominance (VMD) model where a hadron-like photon is treated as a linear combination of vector-mesons states \cite{Gluck:1991jc}. Independently of the chosen approach, one ends up on modelling collisions between a (generalized) hadron and a hadron or nuclear target. 

The modelling of photon-induced processes with general-purpose Monte Carlo event generators has been recently gaining some attention. In \Pythia, a new implementation of such processes has been available for a few years already and recently a similar framework has been set up in Sherpa \cite{Sherpa:2019gpd} and to some extend also in Herwig \cite{Bellm:2015jjp}. In all cases a reasonable agreement with various HERA and LEP data has been found \cite{Hoeche:2023gme, Helenius:2024rth}. However, so far no attempt has been made to model interactions between a real photon and a nuclear target. Such a setup would, however, be very useful for experimental analyses and would provide a realistic baseline for correlation studies and underlying event for jet production. In case of HERA photoproduction data it has been observed that multiparton interactions (MPIs) are a necessary ingredient to describe the measured multiplicity distributions \cite{ZEUS:2021qzg}, which can be naturally explained with additional partonic interactions between the remnant of the resolved photon and the target hadron. In the case of nuclear target, one needs to take an additional step forward and account also for multi-nucleon interactions, where the remnant of the resolved-photon state interacts with the other nucleons in the nuclei. Comparison between the measured multiplicity distributions in photon-proton \cite{CMS:2022doq} and photon-lead \cite{ATLAS:2021jhn} collisions shows that such effects are important and should be included to obtain realistic estimates for particle production.

Another place where hadron-ion collisions occur in nature is in hadronic cascades initiated by high-energy cosmic rays striking our atmosphere \cite{ParticleDataGroup:2022pth}. One open question in cosmic ray physics concerns the production mechanism for ultra-high-energy cosmic rays. Another is the observation that the number of muons measured in these showers does not match the predictions of air shower simulations such as CORSIKA~8~\cite{Engel:2018akg} -- the so-called ``muon puzzle''. Air shower simulations can be divided into one component for simulating interactions between hadrons and gas atoms in the air, and one component for the propagation of the produced hadrons to their next interaction point. For the interactions, CORSIKA uses external models, the most prominent ones being Sibyll 2.3d \cite{Engel:2019dsg}, EPOS-LHC \cite{Pierog:2013ria}, and QGSJet-II.04 \cite{Ostapchenko:2010vb}. With the development of generic hadron-hadron interactions in \Pythia also included a toy model for including nuclear effects, but recent work has shown that it gives predictions that are not fully consistent with the other reigning models \cite{Reininghaus:2023ctx}.

In \Pythia, proton-nucleus and nucleus-nucleus collisions are simulated with the \Angantyr model \cite{Bierlich:2018xfw}. Originally developed as a plugin, this model is now integrated as a part of core program. In case of heavy-ion collisions, the model relies on partonic interactions instead of explicitly including a strongly interacting thermalized medium, i.e. the quark gluon plasma. It has been shown that by including string interactions in hadronization and final-state rescattering, some of the observed collective effects can be reproduced in lead-lead collisions at the LHC \cite{Bierlich:2021poz}. In case of proton-lead collisions, the \Angantyr model provide a good description of the measured multiplicities \cite{Bierlich:2018xfw}.

In this work we introduce an extension of \Angantyr model in \Pythia that accepts any hadron (except multi-heavy flavoured baryons, whose PDFs are not implemented in \Pythia) as a beam particle colliding with a target nucleus. Here our main motivation is to apply this to setup up an explicit VMD model that can be applied to simulate the minimum bias photon-nucleus collisions, and only briefly discuss the application to cosmic rays. For realistic simulations, the shape of photon energy-spectrum needs to be considered which we have here accounted by allowing varying beam energies that can be matched to the calculated photon flux. We also include a feature to switch between different vector mesons on an event-by-event basis. 
We first introduce our simulation framework in Section \ref{sec:modelling}, validate it against HERA data in section \ref{sec:HERAcomp} and present comparisons to UPC data in section \ref{sec:UPCcomp}. We pay special attention to two-particle correlations and perform the template fitting procedure applied by ATLAS to obtain Fourier decomposition with and without non-flow subtraction in Section \ref{sec:vnn}. The work is summarized in section \ref{sec:summary} where also possible future directions are outlined.

\section{Theoretical Background}

\subsection{Generic hadronic interactions}
\label{sec:hadronicInteractions}
Interactions between generic hadrons scattering against a nucleon were implemented in \Pythia~\warning{8.307} \cite{Sjostrand:2021dal}. In this framework, an interaction type is determined randomly according to the partial cross sections of the hadrons involved, whose relative compositions vary from species to species. Once a process type has been selected, a hadron can either scatter elastically, or break up in a diffractive or non-diffractive scattering. In the latter case, the hadron is decomposed into its constituent quarks as determined by its PDFs. After this decomposition, the rest of the process interacts only with the partons, and is agnostic to the original hadron species. Hence, the differences between proton-proton and hadron-proton interactions in \Pythia can be captured by changing the cross sections and the PDFs.

The total cross section can be written as the sum of the partial ones,
\begin{equation}
    \sigtot = \sigel + \sigND + \sigma_{XB} + \sigma_{AY} + \sigma_{XY} + \cdots
\end{equation}
In our approach, we parameterize the total, elastic, and diffractive cross sections, and define the non-diffractive cross section as the remainder, 
\begin{equation}
    \sigND = \sigtot - \sigel - \sigND - \sigma_{XB} + \sigma_{AY} - \sigma_{XY} - \cdots
\end{equation}
In \citeref{Sjostrand:2021dal}, the total cross section is based on the Donnachie-Landshoff model \cite{Donnachie:1992ny},
\begin{equation}
    \sigtot^{AB} = X^{AB} s^\epsilon + Y^{AB} s^{-\eta},
    \label{eq:DLparam}
\end{equation}
where $s$ is the squared CM energy in units of GeV$^2$, the exponents $\epsilon = 0.0808$ and $\eta = 0.4525$ are parameters that do not depend on the hadron in question, and $X^{AB}$ and $Y^{AB}$ are parameters that do depend on the hadron species. The $s^\epsilon$ and $s^{-\eta}$ terms correspond to pomeron and reggeon exchanges, respectively. Going beyond the most common hadrons, such as protons and pions, very little data exists for these cross sections, and the parameters are based on very simple considerations such as taking linear combinations of other hadrons with the same valence content, or applying the additive quark model (AQM) \cite{Lipkin:1973nt,Levin:1965mi}, as described in \citeref{Sjostrand:2021dal}. Diffractive cross sections are calculated using the Schuler and Sjöstrand ansatz \cite{Schuler:1993wr,Schuler:1996en}, with expressions such as 
\begin{equation}
    d\sigSD = \frac{g_{3\Pom} \beta_{A\Pom} \beta_{B\Pom}^2}{16\pi} \frac{dM_X^2}{M_X^2} (e^{B_{XB} t} dt) \, F_\mathrm{SD}(M_X^2, s) 
\end{equation}
for single diffraction $AB \to XB$. Here, the hadron species-dependent parameters are the Pomeron couplings $\beta$, which with the appropriate ansatz and normalization can be expressed in terms of $X^{A\mathrm{p}}$, and the exponential slope $B_{XB}$, which depends on whether hadron $B$ is a baryon or meson. Similar expressions are used for the elastic and double diffractive components,
\begin{equation}
    d\sigel = (1 + \rho^2) \frac{\sigtot^2}{16\pi} e^{B_\mathrm{el} t} dt,
\end{equation}
\begin{equation}
    d\sigma_{XY} = \frac{g_{3\Pom} \beta_{A\Pom} \beta_{B\Pom}}{16\pi} \frac{dM_X^2 dM_Y^2}{M_X^2 M_Y^2} (e^{B_{XY} t} dt) \, F_\mathrm{DD}(M_X^2, M_Y^2, s).
\end{equation}
The non-diffractive cross section is given in the end as the difference between the total and the other partial cross sections.

There is very little data for PDFs beyond nucleons, and as such, the models used in \Pythia are very simple. The basic ansatz is that of Gl\"uck, Reya et al. \cite{Gluck:1991ey,Gluck:1999xe}, which takes the valence, sea and gluon distributions at some initial scale $Q^2_0$ to be on the form
\begin{equation}
    f(x, Q_0^2) = N x^a (1-x)^b (1 + A \sqrt{X} + B X),
\end{equation}
which can then be evolved to higher scales through DGLAP evolution. In \Pythia, this is simplified further for the valence content by assuming $A = B = 0$. There is no solid theory for determining $a$ and $b$, but one guiding principle is that all valence quarks must have roughly the same velocity in order for the hadron to stay intact, and therefore heavier quarks must have a smaller $b$ and a larger momentum fraction $\langle x \rangle$. These values are chosen for each hadron based on heuristic guesses, and the normalization factor is given by the valence sum relations. The gluon and u/d sea distributions at the initial scale are assumed to be the same as for the pion, except multiplied by a factor $x^d$ in order to soften them in light of the increased $\langle x \rangle$. Finally, the gluon and sea distributions are scaled by a normalization factor in order to satisfy the momentum sum relation.

\subsection{\Angantyr}

\Angantyr is the heavy ion model of \Pythia \cite{Bierlich:2018xfw}. The core idea of this model is that an interaction between two nuclei can be described as a number of nucleon-nucleon interactions stacked on top of each other, with a special treatment of wounded nucleons that have already interacted, where the individual interactions can be simulated by \Pythia's existing minimum-bias machinery.

Event generation in \Angantyr begins with a Glauber modelling of subcollisions \cite{Glauber:1955qq}. First, the spatial distribution of nucleons in each beam is determined, by default using the Glissando model \cite{Rybczynski:2013yba}, and the beam--beam impact parameter $b_{AA}$ picked. By default, $b_{AA}$ is sampled according to a Gaussian, which gives a bias towards lower values that is compensated by reweighting. This is in order to improve the statistics for central collisions at the expense of precision at peripheral collisions. 

Knowing the impact parameter of the collision, $b_{AA}$, and the nucleon configurations of each beam, one can derive the nucleon-nucleon impact parameter $b_{NN}$ for each nucleon pair.
For each such pair, a potential subcollision type is chosen. The relative probabilities are functions of $b_{NN}$ and the hadronic partial cross sections from \secref{sec:hadronicInteractions}. The exact dependence on the impact parameter is not well known, and \Angantyr offers a few options. The most simplistic model is a black disk with a sharp division between the region for each type, starting with absorptive being the most central (here, ``absorptive'' basically means non-diffractive, except it may lead to diffractive topologies if either nucleon is wounded). That is, the collision is labelled as absorptive if $\pi b^2 \leq \sigabs$, double diffractive if $\sigabs < \pi b^2 \leq \sigDD$, and so on, continuing with single diffractive, and elastic scatterings being the most peripheral. If $\sigtot < \pi b^2$, the nucleon pair is labelled as not interacting.

A more sophisticated collision model, which is the default in \Angantyr, is based on an approach by Strikman et al. \cite{Heiselberg:1991is,Blaettel:1993ah,Alvioli:2013vk,Alvioli:2014sba,Alvioli:2014eda}, wherein the radius $r$ of each nucleon is allowed to fluctuate. In proton-proton collisions, the sampling of the overlap is based on a parametrization fitted to data and thus already accounts for fluctuations. In heavy ion collisions, however, fluctuations have larger impact: when the projectile proton fluctuates to a large radius, it leads to more sub-collisions, which in turn leads to a significantly longer tail in the multiplicity distribution that is difficult to model without such fluctuations. Especially in $pA$ collisions, fluctuations in the projectile are more significant than fluctuations of nucleons in the target nucleus. Similarly to the black disk approach, more central collisions are more likely to be absorptive, but unlike the black disk, the division is continuous. 

Specifically, the radius $r$ of each nucleon follows a gamma distribution with mean radius $r_0$ and shape parameter $k_0$. As the ``size'' of the nucleons is what determines the cross section, these parameters can be related to the total cross section according to
\begin{equation} \label{eq:symmetricParametersRelation}
\sigtot/\pi = 4 (k_0 + k_0^2) r_0^2,
\end{equation}
and as the total cross section is given by \Pythia's hadronic model, we treat $r_0$ as a dependent parameter. After sampling the projectile- and target radii $r_p$ and $r_t$, the \emph{opacity} is calculated as
\begin{equation}
    T_0 = \left[ 1 - \exp\left(-\frac{\pi (r_p + r_t)^2}{\sigma_d} \right) \right]^\alpha,
\end{equation}
introducing two new parameters, $\sigma_d$ and $\alpha$, for a total of three free parameters overall. A given set of these parameters fixes the total and partial cross sections, which are already known from the hadronic collision model, constraining the possible values of the parameters. In \Angantyr, the parameters are determined by a genetic algorithm. It begins by choosing random parameter sets, and selects the set that gives the best fit to the cross sections. Next, it produces random variations of this set, and again selects the best fit. This process is iterated for several generations (20 by default), eventually converging to a parameter set that accurately reproduces the cross sections. In addition, the elastic $b$-slope (appearing in $d\sigel/dt \sim e^{bt}$) is used as a fitting target in this process.

After all subcollision candidates have been found they are sorted according to type, starting with absorptive collisions, followed by double diffractive, single diffractive, and finally elastic collisions. Within each process type, the collisions are sorted according to $b_{NN}$, starting with the most central. The collisions are then simulated in order, using \Pythia to generate interactions to parton level. If a particle participates in several subcollisions, the subsequent ones are handled using the wounded nucleon model. When a wounded projectile particle collides with a target, the projectile emits a pomeron-like particle, which imparts some momentum on the beam remnants from the original collision. The secondary collision is then treated as a non-diffractive collision between the pomeron-like particle and the target, leading to a diffractive topology between the projectile and target.

The impact parameter $b_{NN}$ determines the probability of each type of process, but $b_{NN}$ also plays a different role in event generation, namely in determining MPI activity: in non-diffractive events, a smaller $b_{NN}$ is correlated with a larger number of parton interactions. However, the different subcollision models may give different $b_{NN}$ distributions within each type of collisions. For example, for a black disk where $b_{NN}$ is assumed to be distributed uniformly on a disk with radius $\sqrt{\sigabs/\pi}$, the mean radius is $\langle b_\mathrm{ND} \rangle = \frac23 \sqrt{\sigabs/\pi}$, whereas a fluctuating model could lead to an increased number of collisions with large impact parameter resulting as an increased $\langle b_\mathrm{ND} \rangle$ compared to black-disk one.

The MPI framework in \Pythia \cite{Sjostrand:2004pf} does not use the physical impact parameter values but the overlap $\mathcal{O}(b)$ is parametrized in terms of scaled impact parameter $b_{\mathrm{MPI}}$ for which $\langle b_{\mathrm{ND}}^{\mathrm{MPI}}\rangle = 1$. Thus it is necessary to scale the $b_{NN}$ values with $\langle b_{\mathrm{ND}} \rangle$ when passing this information to \Pythia for MPI generation for a given nucleon-nucleon collision. The scaling is calculated as $b_\mathrm{MPI} = f \cdot b_\mathrm{NN} / \langle b_\mathrm{ND} \rangle$ where parameter $f$ (0.85 by default) can be used to compensate for potentially different $b$ distributions between \Angantyr and the MPI framework in \Pythia. 

Once all subcollisions have been resolved this way, the resulting partons from each event are all combined into one overall event. Partons from different subevents may become correlated through colour reconnections \cite{Sjostrand:2004pf}, and optional features such as rope formation \cite{Bierlich:2016vgw} and string shoving \cite{Bierlich:2017vhg,Bierlich:2020naj} may be simulated at this point. Finally, the partons are hadronized and post-hadronization effects such as decays and rescattering are simulated.

\subsection{Hadronic structure of quasi-real photons}

One of the main motivations of developing the generic hadron-ion collision framework is that it allows to simulate an important aspect of the photon-nucleus collisions. As in the considered processes the photons have low virtuality, they may fluctuate into a hadronic state with equal quantum numbers. Thus to have a complete picture of these photon-induced processes, in addition to interactions with direct photons, also the component where photon has a resolved partonic structure has to be accounted for. The resolved photons consists of a component of perturbative origin, which can be directly calculated, and from a non-perturbative transition to a hadronic state. The latter component can be modelled with the VMD where the hadronic state is given by a linear combination of different vector mesons. While the direct photons have a significant contribution for high-$p_{\mathrm{T}}$ observables, such as jets, it turns out that the bulk of the cross section is actually arising from collisions with resolved photons. Furthermore, as these collisions are essentially just like any other hadronic collisions, there can be several partonic interactions at the same collisions. A full-fledged framework for photoproduction with proton target is been included in \Pythia 8.3 and described in detail in Ref.~\cite{Bierlich:2022pfr} and some applications considered in Refs.~\cite{Helenius:2017aqz, Helenius:2019gbd}.

In this work we focus on collisions where the interactions of quasi-real photons and protons or heavy nuclei generate a large number of final state particles. The production of these high-multiplicity events is dominated by the hadron-like part for the reasons explained above. Therefore, instead of including all components of real photons on one go, we implement an explicit VMD model and model real photons as a combination of hadronic states. Following Ref.~\cite{Schuler:1992dt}, the total cross section for this contribution in photon-proton collision at a collision energy $s$ is given by
\begin{equation}
\sigma_{\mathrm{VMD}}^{\gamma \mathrm{p}}(s) = \sum_{V} \frac{4 \pi \alpha_{\mathrm{em}}}{f^2_V} \sigma^{V \mathrm{p}}(s),
\label{eq:vmd}
\end{equation}
where the coupling factor $4 \pi \alpha_{\mathrm{em}}/f^2_V$ is the probability for a photon to fluctuate into a vector-meson state $V$ and the $\sigma^{V \mathrm{p}}(s)$ the cross section for vector-meson-proton interaction. We use $f_V$ values from Ref.~\cite{Bauer:1977iq} where $f_{\rho} = 2.2$, $f_{\omega} = 23.6$, $f_{\phi} = 18.4$, $f_{J/\psi} = 11.5$ and cross section parametrizations follow the form in equation~(\ref{eq:DLparam}) with the state-dependent parameter values from Ref.~\cite{Schuler:1994ft}. The vector mesons state participating to the interaction is sampled according to the relative contributions of different vector-meson state to the total cross section at a given collision energy as given in equation~(\ref{eq:vmd}).

\subsection{Photon fluxes from different beams}

All charged particles accelerated to high energies may emit photons but the energy spectrum and possible virtualities depend on the structure and size of the particle. Within the equivalent photon approximation (EPA) \cite{Budnev:1975poe}, derived in the limit $Q^2 \rightarrow 0$, the photon flux can be factorized from the hard scattering. In case of point-like charged leptons the photon flux, $f_{\gamma}$, can be calculated from the well known Weizsäcker-Williams formula \cite{vonWeizsacker:1934nji, Williams:1934ad}
\begin{equation}
f_{\gamma}^{l}(x,Q^2) = \frac{\alpha_{\text{em}}}{2 \pi} \frac{1 + (1-x)^2}{x} \frac{1}{Q^2},
\label{eq:photonFluxEdiff}
\end{equation}
where $x$ is the momentum fraction carried by the photon wrt. the beam particle emitting the photon and $Q^2$ the (space-like) virtuality of the photon. Photoproduction is usually separated from deep-inelastic scattering events by requiring an upper limit for the virtuality $Q^2_{\text{max}} = \mathcal{O}(\text{GeV})$ to ensure that the EPA is applicable. The lower limit for the virtuality follows from kinematics and is proportional to the lepton mass, see details for the \Pythia implementation in Ref.~\cite{Bierlich:2022pfr}.

In case of charged hadrons and nuclei, one needs to take into account the finite size of a particle emitting the photon by including a form factor into the flux calculation. The form factor ensures that the emitting particle stays intact and suppresses the virtuality of the photons. In case of protons the photon flux rapidly vanishes beyond $Q^2 > 1~\text{GeV}^2$ whereas with heavy ions the virtualities are restricted to some tens of MeVs \cite{Bertulani:2005ru}. In case of heavy ions, a good approximation of the photon flux in the impact parameter space $b$ can be calculated by assuming a point-like charge and rejecting interactions with small enough impact parameter where also strong interactions would be possible. Usually the cut in impact parameter space $b_{\text{min}}$ is taken to be around $2 R_A$, where $R_A$ is the radius of the colliding nuclei $A$. Integrating the flux over the allowed impact-parameter space give the flux as
\begin{equation}
f_{\gamma}^{A}(x) = \frac{\alpha_{\text{em}} Z^2}{\pi x} \left[ 2 \xi K_1(\xi) K_0(\xi) - 
  \xi^2(K^2_1(\xi)-K^2_0(\xi))\right],
\label{eq:photonFluxAint}
\end{equation}
where $Z$ is the charge of the nucleus, $K_i$ modified Bessel functions of the second kind and $\xi = b_{\text{min}} x m_{N}$, $m_{N}$ being the per-nucleon mass. In the presented results for UPCs in the Pb-Pb collisions at the LHC we use the flux in equation (\ref{eq:photonFluxAint}) with $b_{\mathrm{min}} = 13.272~\text{fm}$, $m_{N} = 0.9314~\text{GeV}$ and $Z=82$.

\section{Modelling hadron-ion interactions}
\label{sec:modelling}

\subsection{Subcollisions with generic hadrons}
\label{sec:hadron-ion}

Our goal is now to extend \Angantyr and implement $hA$ collisions for generic projectile hadrons $h$. To this end, many of the components we need already exist in \Pythia/\Angantyr. In fact, once the subcollisions are all determined, the hadronic interactions are fully implemented in \Pythia. Furthermore, in the wounded nucleon model, the momentum imparted by the pomeron-like particle on the beam remnants of the wounded hadron follows the same distribution regardless of the species of the original beam hadron \cite{Bierlich:2018xfw}. After subcollisions have been generated, we are left with a partonic state that is also agnostic to the original hadrons, and the existing framework is completely applicable after this point. The only unresolved question is then how to determine subcollisions for different projectile hadrons.

As all partial cross sections are given by \Pythia, in the black disk model we can simply apply the sharp thresholds in the impact parameter to determine each collision type. Complications arise in the fluctuating model. While it is possible to assume only three free parameters as before, the projectile and target hadrons may now be of different species, and thus can have different parameters for fluctuations. Assuming the projectile/target each follow a gamma distribution with mean radius $r_{p/t,0}$ and shape parameter $k_{p/t,0}$, the relation in \eqref{eq:symmetricParametersRelation} instead becomes
\begin{equation}
    \frac{\sigtot}{\pi} = (k_{p,0} + k_{p,0}^2) r_{p,0}^2 + (k_{t,0} + k_{t,0}^2) r_{t,0}^2 + 2 k_{p,0} k_{t,0} r_{p,0} r_{t,0}.
\end{equation}
Although it is possible in principle to solve this for one of the parameters, say $r_{t,0}$, it is a second degree equation, which means that real solutions only exist in some regions of parameter space. For the sake of simplicity, we ignore this relation and simply work with a rectilinear parameter space with six independent parameters, $k_p$, $r_p$, $k_t$, $r_t$, $\sigma_d$, and $\alpha$. The drawback of this is that with more parameters, the number of iterations required for a converged fit in the evolutionary algorithm increases significantly.

\begin{figure}
    \centering
    \includegraphics[width=0.45\textwidth]{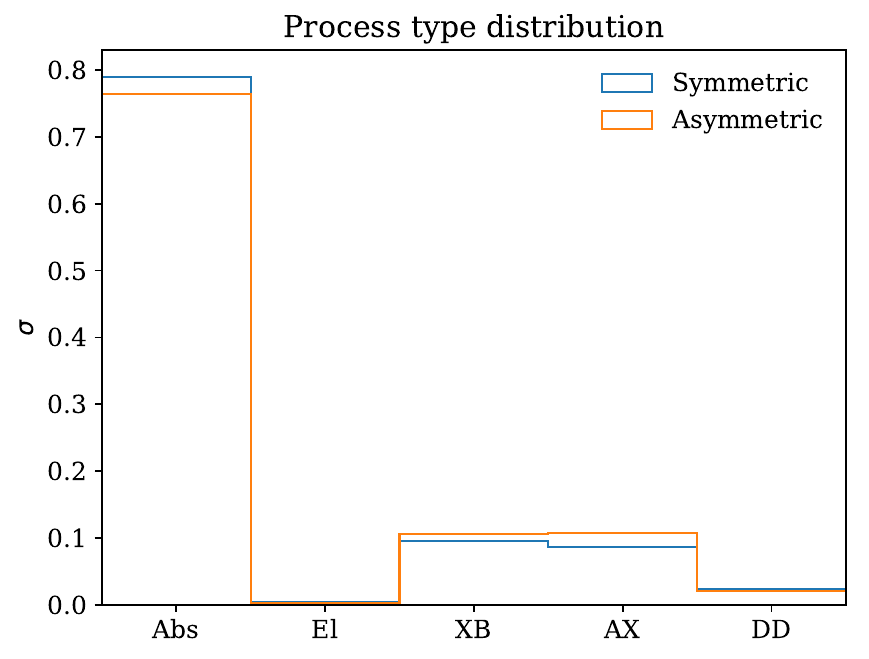}
    \includegraphics[width=0.45\textwidth]{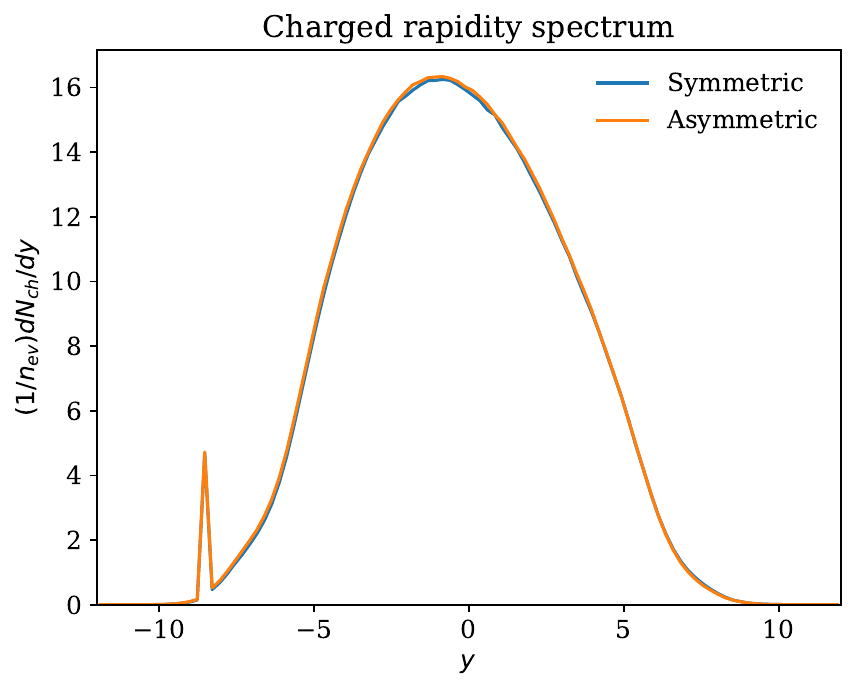}
    \caption{Comparison between model with symmetric and asymmetric parameters for $\pi^0 p$ collisions at 5.02~TeV. a) Relative composition of event types (absorptive, elastic, projectile diffraction (XB), target diffraction (AX) and double diffraction). b) Charged rapidity spectra.}
    \label{fig:fluctuations}
\end{figure}

In \figref{fig:fluctuations}, we evaluate the effect of the six-parameter model compared to one where the two sides are assumed to have the same parameters, in the case of $\pi^0\mathrm{Pb}$ collisions at $5.02$~TeV. After training, the goodness of fit was $\chi^2/dof = 2.6$ for the symmetric case and $6.4$ for the asymmetric case, despite the fact that for the symmetric case, the model was trained for 30 generations, while the asymmetric case needed more than 50,000 generations to achieve this level of precision. In the asymmetric case, we observed that the precision plateaued at several times during the training, indicating that the model found and stayed at a local minimum in parameter space until random variations found a better point. Considering \figref{fig:fluctuations}a, one advantage of the second model is that we see an asymmetry between the rate of diffraction in the projectile and target. This difference cannot be captured by the symmetric model because it gives equal cross sections for diffraction on each side, even when the target cross sections are different. However, the difference here is small enough compared to the overall uncertainty in the cross sections that it has little practical value. Indeed, as we see in \figref{fig:fluctuations}b), the actual observables are virtually indistinguishable.

In order to properly handle the effects of independent fluctuations, one would have to develop the underlying model further and choose fitting targets that are sensitive to the individual fluctuations, e.g. the expected mean radius of each hadron. In the current version, given the similarity of the results between the symmetric and asymmetric models and the large computationally overhead in the latter, we conclude that there is no clear reason to favour the asymmetric model over the symmetric one. 

\subsection{Model tests}

\begin{figure}
    \centering
    
    \subfloat[]{\includegraphics[width=0.45\textwidth]{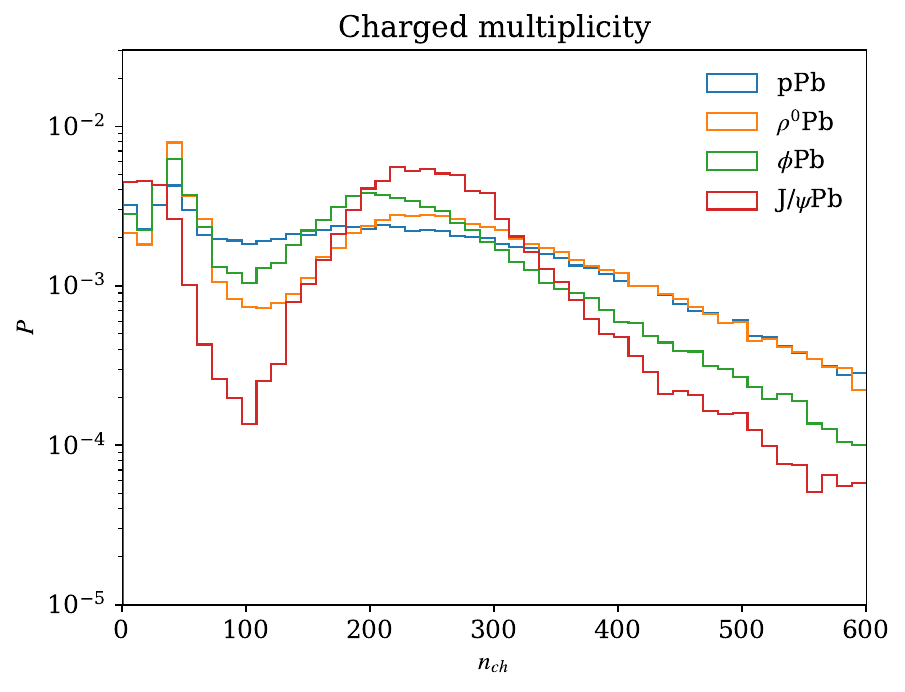}}
    \subfloat[]{\includegraphics[width=0.45\textwidth]{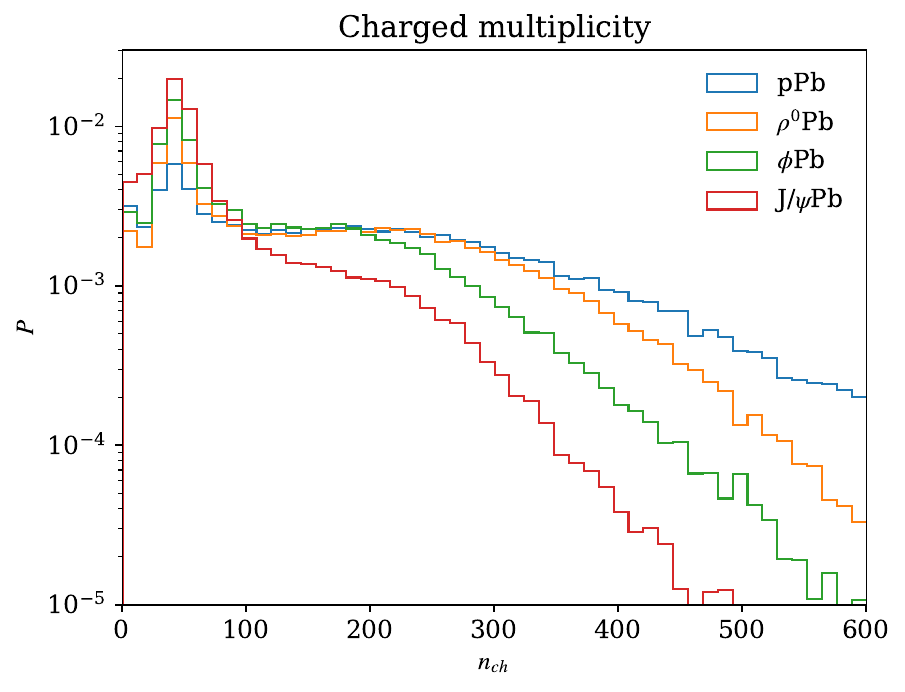}}
    \\
    \subfloat[]{\includegraphics[width=0.45\textwidth]{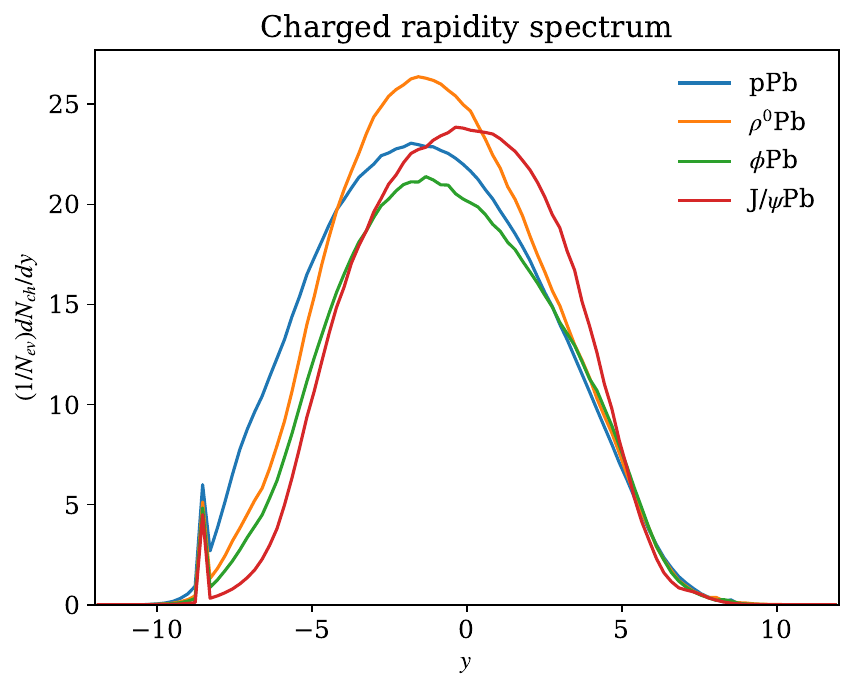}}
    \subfloat[]{\includegraphics[width=0.45\textwidth]{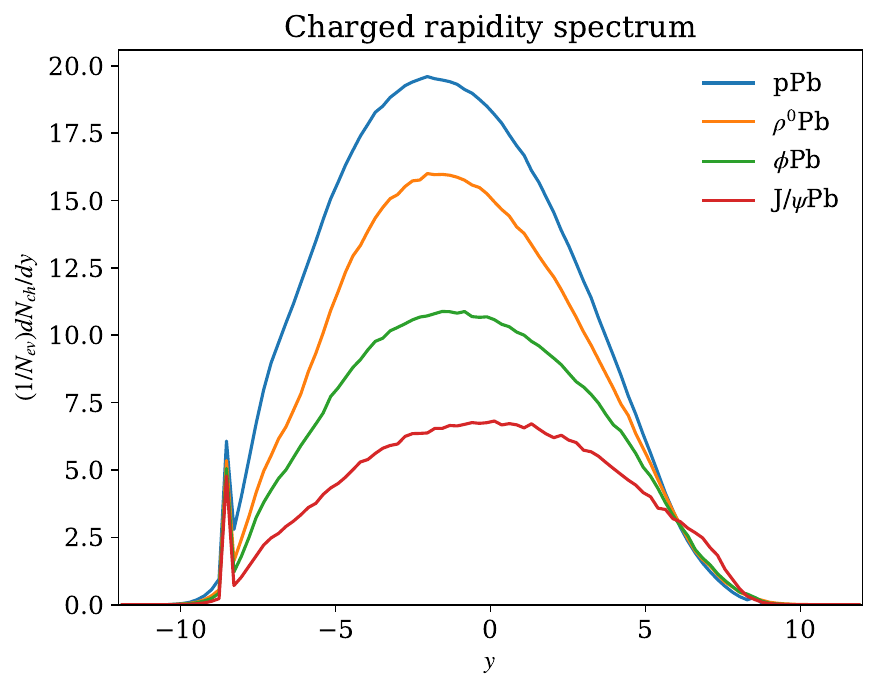}}
    \caption{Comparison of observables for various hadron-proton and hadron-lead beam combinations at 5.02~TeV. The left-hand figures correspond to an $\langle b_\mathrm{ND} \rangle$ calculated by the fluctuating model, while the figures on the right use the black disk $\langle b_\mathrm{ND} \rangle$. The spikes around $y = -8$ in the rapidity spectra correspond to the nuclear remnant, constructed as a combination of spectator nucleons.}
    \label{fig:observables}
\end{figure}

As a basic model test, the charged multiplicity and rapidity distributions for different $h \Pb$ combinations are shown in \figref{fig:observables}. In the charged multiplicities with default parameters (\figref{fig:observables}a), we see a bimodal distribution, particularly prominent in the $J/\psi$ case. Furthermore, we see a convoluted progression where the $\phi$ peak is below both $\rho^0$ and $J/\psi$, even though one might expect a hierarchical relationship between $\rho^0$, $\phi$ and $J/\psi$ as they contain progressively heavier $q\bar{q}$ pairs. The reason for these behaviours is that with the fluctuating model, hadrons like $J/\psi$ would sometimes fluctuate to very large radii, resulting in an unnaturally large $\langle b_\mathrm{ND} \rangle$. In turn, this gives a very small $b_\mathrm{MPI}$, basically treating all collisions as head-on for the purposes of calculating MPI activity, and all absorptive collisions are pushed towards higher multiplicities, giving a prominent second peak. 

In order to circumvent this issue, we offer an alternative where the average impact parameter is calculated under the black disk assumption, 
\begin{equation}\label{eq:avNDb}
\langle b_\mathrm{ND} \rangle = \frac23 \sqrt{\sigabs/\pi},
\end{equation}
but still allowing hadron size to fluctuate and maintain some of the multiplicity tail. The resulting multiplicities, shown in \figref{fig:observables}b, have a smoother behaviour and a clear hierarchy between $\rho^0$, $\phi$ and $J/\psi$. Applying the $\langle b_\mathrm{ND} \rangle$ from the black-disk approximation could potentially underestimate the result impact-parameter values and the variation between the two options could be to some extent taken as a modelling uncertainty.

In the rapidity spectra, shown in \figref{fig:observables}c and d, we see more clearly how the two different models affect the average multiplicity. For $J/\psi$, the effect is massive, with the two differing by a factor of roughly 4. For $\rho^0$ and $\phi$ the difference is around a factor 2, which is still large, but in line with other model uncertainties such as PDFs. The rapidity spectra also show a shift due to additional subcollisions, especially clear in the proton case where one would normally expect $pp$ rapidity spectra to be symmetric. For the heavier $\phi$ and $J/\psi$, there are fewer subcollisions and the projectile itself is heavier, and therefore lead to a lesser shift, as expected. 

\subsection{Cross sections}

Cross sections for various hadron-lead combinations at 5.02~TeV are shown in \tabref{tab:crossSections}, together with the corresponding hadron-proton cross sections from the DL/SaS models. As a first na\"ive estimate, one might expect that the $hA$ cross section should equal the $hN$ cross section multiplied by the nucleon number $A$ of the ion, e.g. $A = 208$ for lead. In practice, however, the target nucleons are bunched up, causing a shadowing effect. That is, in a central collision where the projectile has several chances to interact with nucleons in the target, a collision is almost guaranteed. Hence, we see that even though the $J/\psi \p$ cross section is almost an order of magnitude smaller than the $\p\p$ one, the corresponding nuclear cross sections differ by less than a factor of 2.

\begin{figure}
    \centering

    \subfloat[]{\includegraphics[width=0.49\linewidth]{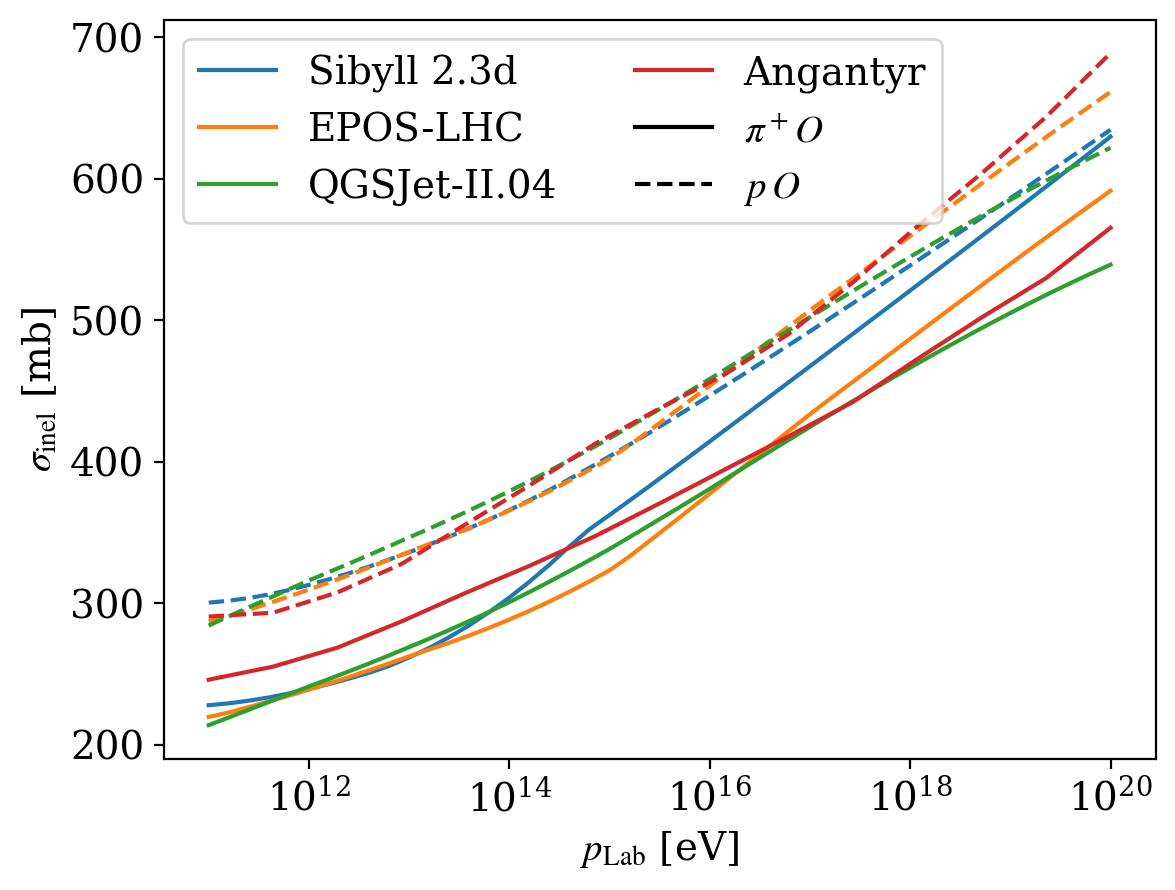}}
    \subfloat[]{\includegraphics[width=0.49\linewidth]{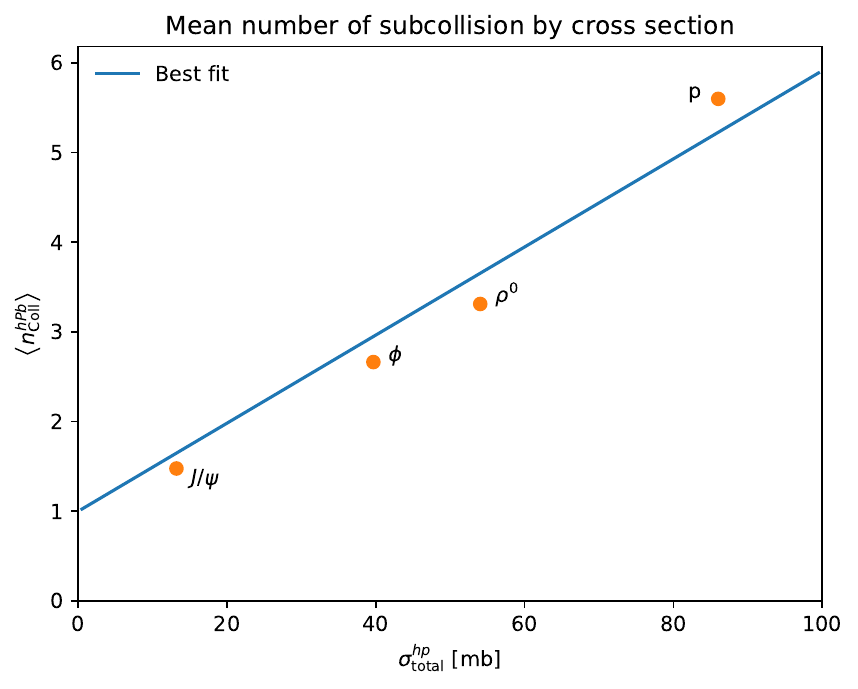}}
    \caption{a) A comparison of \Angantyr inelastic cross sections to corresponding cross sections from other models \cite{Reininghaus:2023ctx}, ranging from $p_\mathrm{Lab} = 10^{11}$ eV ($\eCM \sim 13$~GeV) to $10^{20}$ eV ($\eCM \sim 250$~TeV). b) Average number of subcollisions in $h\Pb$ collisions at 5.02~TeV. The best fit is assumed to be on the form $1 + k\sigtot$}
    \label{fig:crossSections}
\end{figure}

Nuclear cross sections with an Oxygen target are shown in \figref{fig:crossSections}a for $\pi^+ \mathrm{O}$ and $\mathrm{p}\mathrm{O}$ as a function of $p_\mathrm{Lab}$, comparing to the values calculated by Sibyll 2.3d \cite{Engel:2019dsg}, EPOS-LHC \cite{Pierog:2013ria}, and QGSJet-II.04 \cite{Ostapchenko:2010vb}. Whereas a comparison with the earlier toy model in \Pythia gave results that were inconsistent with the other models \cite{Reininghaus:2023ctx}, we now see a much closer match. Note that in \Pythia~8.312 there is a numerical issue in \Angantyr where event generation fails at the highest lab momenta in \figref{fig:crossSections}a, which we circumvented by calculating the cross section in the corresponding CM frame, where the numerical values are smaller and boosts more numerically stable.

At the lowest energies, \Angantyr gives a higher $\pi \mathrm{O}$ cross section than the other models, but the discrepancy is overall quite small. This cross section then grows noticeably more slowly than in the other models, in accordance with how the $\pi^+\mathrm{p}$ cross section in \Pythia grows more slowly than the other models \cite{Reininghaus:2023ctx}. A similar trend is seen for $\mathrm{p O}$, but here the difference between the hadronic cross sections is smaller, suggesting there may be other model effects at play. However, since the difference is very small, we have not investigated this further.

Once the nuclear configuration and impact parameter have been determined and it has been decided that at least one $hN$ collision occurs, the number of additional nucleons within range scales roughly with the total hadronic cross section (in the case of a uniform target, the scaling would be exact). Therefore, we expect that the mean number of subcollisions roughly follows the shape $\langle \ncoll \rangle = 1 + k \sigtot$, where $k$ represents the density of nucleons in the projection of the nucleus onto the $xy$-plane. This is shown in \figref{fig:crossSections}b where a line with a fitted value for $k$ capture the trend of the calculated points well.

\begin{table}[htb]
    \centering
    \begin{tabular}{c c c c c c}
        $h$ &  $\sigtot^{h\mathrm{Pb}}$ & $\sigtot^{h\mathrm{p}}$ & $\sigND^{h\mathrm{p}}$ & $\sigdiff^{h\mathrm{p}}$ & $\sigel^{h\mathrm{p}}$ \\
        \hline
        $p$      & $2.05 \times 10^3$ & 86.0 &  47.7 & 19.9 & 18.4 \\
        $\rho^0$ & $1.95 \times 10^3$ & 54.0 &  33.5 & 12.5 &  8.0 \\
        $\phi$   & $1.81 \times 10^3$ & 39.7 &  26.9 & 8.5 &  4.3 \\
        $J/\psi$ & $1.26 \times 10^3$ & 13.2 &  10.5 & 2.2 &  0.5 
    \end{tabular}
    \caption{Hadron--proton and hadron--lead cross sections for different hadron species at 5.02~TeV, in units of mb. The first column corresponds to $h$Pb, and subsequent columns are $h$p. Note that $hA$ cross sections cannot easily be calculated \emph{a priori} in \Angantyr, and were calculated using Monte-Carlo integration.}
    \label{tab:crossSections}
\end{table}

\subsection{Variable energies}
In photon-induced VMD processes, the available energy in the hard interaction changes on an event-by-event basis, depending on the photon flux. The energy also varies in hadronic cascades, with less energy further down in the cascade. Thus, there is a technical need for efficiently changing the energy in each event. Moreover, there is similarly a need to switch between different beam particles. In VMD processes, one needs to switch between different vector mesons, while in hadronic cascades, one must include all hadrons that are long-lived enough to rescatter in the medium.

This feature is implemented in \Pythia hadron-hadron. It requires a number of technical steps in the code, most significantly handling energy-dependent MPI parameters. These parameters are calculated at the beginning of the run and constitute the slowest part of initialization, which makes it unfeasible to perform this calculation every time the beam energies change. \Pythia solves this by calculating the parameters at several energy values during initialization, then using interpolation to determine values at a given energy during event generation. 

Implementing variable energies in \Angantyr is the same story. Here, the bottleneck is the evolutionary algorithm used to determine nuclear fluctuations. We again solve it by calculating the subcollision parameter values during initialization and interpolating during event generation. As an additional optimization, we assume the parameters vary relatively slowly with energy, so that by using the parameters at each interpolation point as the initial values when running the algorithm at the next point, we can achieve a slightly faster convergence. To also include beam switching, the framework automatically handles initializing for different hadron types, saving the initialization data to one file for easy reuse.

\begin{figure}
    \centering
    \subfloat[]{\includegraphics[width=0.4\textwidth]{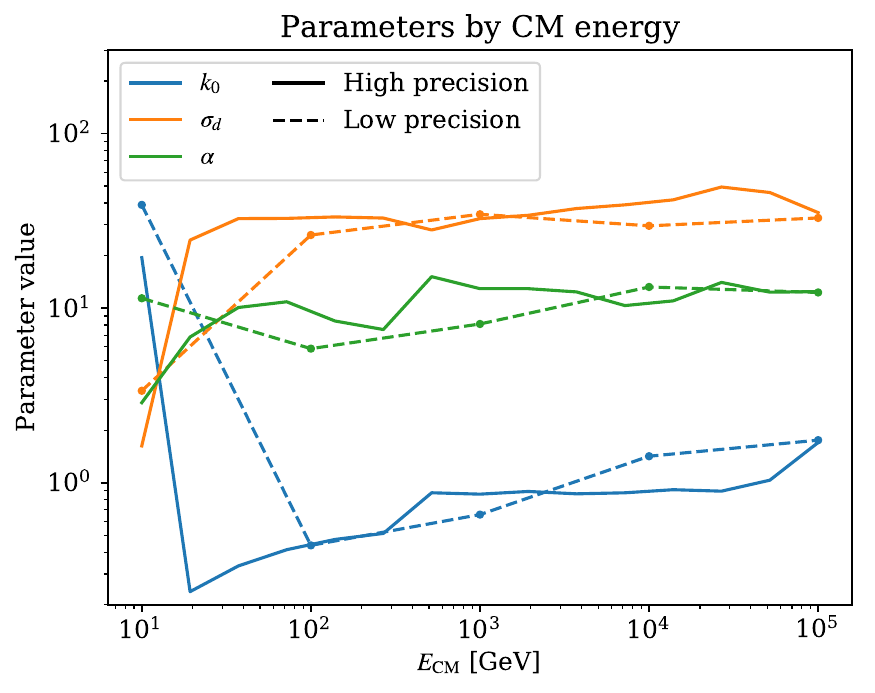}}
    \subfloat[]{\includegraphics[width=0.4\linewidth]{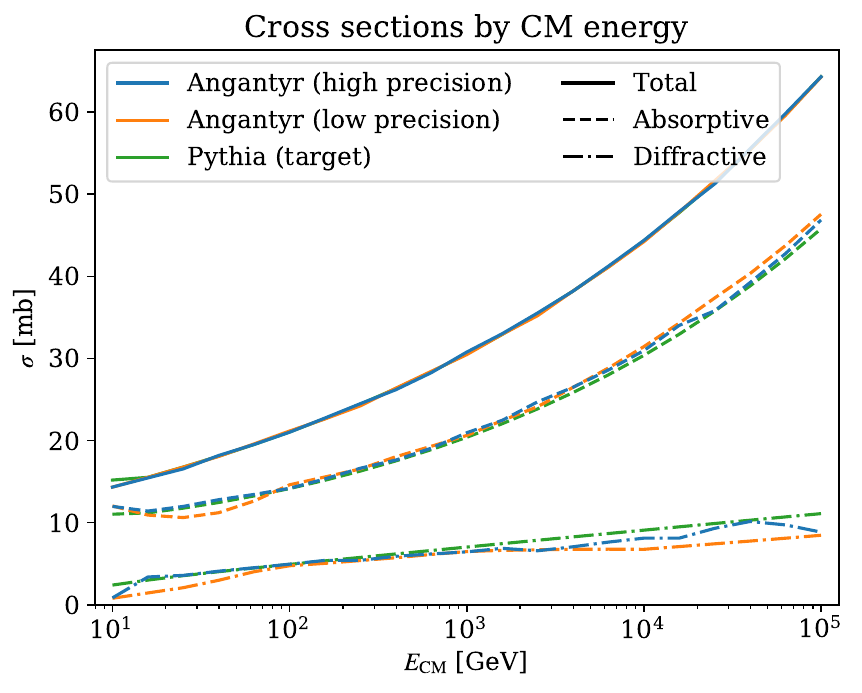}}
    \\
    \subfloat[]{\includegraphics[width=0.4\linewidth]{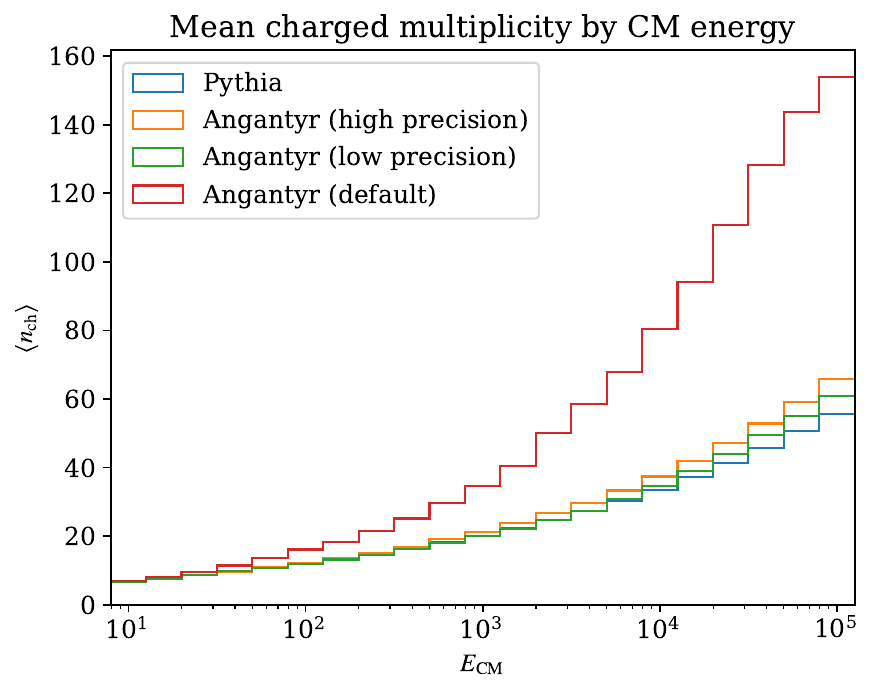}}
    \caption{Comparison of a low-precision cross section fit to one with much higher precision for $\phi \p$ collisions.
    a) Parameters output from the genetic algorithm; b) resulting cross sections, compared to target cross sections; c) mean charged multiplicity in $\phi \p$ collisions as a function of energy, comparing different model choices. Here ``default'' refers to using $\langle b_\mathrm{ND} \rangle$ calculated from fluctuations, while the other \Angantyr plots use the black disk $\langle b_\mathrm{ND} \rangle$.}
    \label{fig:genetic-algorithm}
\end{figure}

The question is then how many generations and interpolation points are necessary. We investigate this in \figref{fig:genetic-algorithm}, where we consider $\phi \p$ collisions at energies ranging from $\eCM = 10^5$~GeV down to $10$~GeV, comparing a case with 5 interpolation points to one with 21 points (recall that the parameters are only a function of the hadron species, i.e. the parameters will be identical for all $\phi A$). In \figref{fig:genetic-algorithm}a, we do see a noticeable discrepancy between the parameter values. However, the fitting procedure might find different local minima in parameter space (as indicated by point-by-point fluctuations in the high-precision fitting), and this discrepancy in itself is not cause for concern. What matters is the resulting cross sections and observables. The cross sections in \figref{fig:genetic-algorithm}b show that the two models give similar results, with the high-precision run giving a marginally better fit to the target.

\figref{fig:genetic-algorithm}c shows the mean charged multiplicities as a function of the CM energy. Here, the high- and low-precision runs use the black disk $\langle b_\mathrm{ND} \rangle$ as described in equation (\ref{eq:avNDb}), and we compare to the case where $\langle b_\mathrm{ND} \rangle$ is calculated from the fluctuations (labelled ``\Angantyr (default)''). First, we see that the high- and low-precison runs both are close to the \Pythia target, and in fact the low-precision version gives a closer match. We have also validated this for $\rho^0$ and $J/\psi$, and conclude that one interpolation point per decade gives a precision that is well within other model uncertainties. With the fluctuating $\langle b_\mathrm{ND} \rangle$, the discrepancy grows large, especially at higher energies relevant for cosmic ray applications. We therefore recommend the black disk $\langle b_\mathrm{ND} \rangle$ for such use cases until a more generic model for hadronic fluctuations capable to handle these very high energies is developed.

In the preceding studies, we have gone all the way down to $10$~GeV where we expect to see non-perturbative effects such as resonance formation. While default \Pythia can model hadronic interactions at low energies, it is not immediately clear how to extend this to heavy ions. In particular, all low-energy processes implemented in \Pythia have purely hadronic final states with no beam remnants, so it is not straightforward to model secondary subcollisions with the wounded nucleon model. Furthermore, a sudden variance in the parameters at low energies in \figref{fig:genetic-algorithm}a indicates that the genetic algorithm might not be as accurate in this region. As a brute force approximation, one might simply assume that hadrons wounded in non-perturbative interactions cannot participate in further collisions. While this would not necessarily be an incorrect physical picture, we note that such a hadron-ion collision would be indistinguishable from a hadron-nucleon collision, except for the addition of nuclear remnants.

We have not implemented a way to do this automatically in \Angantyr, but a user who wishes to do this can work around it by initializing another \Pythia object that is configured to generate only low energy events, and using that object to generate a fraction of the events. Specifically, in \Pythia this fraction is given by 
$P_\mathrm{nonpert} = \exp \left(-\frac{\eCM - E_\mathrm{min}}{E_\mathrm{width}} \right)$, where the parameters are by default $E_\mathrm{min} = E_\mathrm{width} = 10$~GeV. This gives a fraction of around 2~\% at 50~GeV, above which we consider low-energy processes negligible.

\section{Applications}

\subsection{ZEUS data for high-multiplicity events in photoproduction}
\label{sec:HERAcomp}

A recent ZEUS analysis studied azimuthal correlations in high-multiplicity events in DIS and photoproduction \cite{ZEUS:2021qzg} in electron-proton collisions at HERA. In addition to two- and four-particle correlations, the analysis presented results for charged-particle multiplicity and rapidity distributions in photoproduction with a proton target that can be used to validate the vector-meson-dominance model considered in this work. In the analysis, the experimental data were compared to \Pythia simulations for photoproduction by varying the value of $p_{\mathrm{T,0}}^{\mathrm{ref}}$ that determines the probability for MPIs in case of resolved photons and found that values around $3 < p_{\mathrm{T,0}}^{\mathrm{ref}} < 4~\text{GeV}$ were preferred. This maps to slighly lower MPI probability wrt. what has been found optimal in proton-proton collisions \cite{Skands:2014pea} but also the considered energy range is quite different. In any case, the data clearly disfavored results where MPIs were disabled and is in line what has been obtained from a comparison to charged-particle $p_{\mathrm{T}}$ spectra in $\gamma \gamma$ collisions at LEP \cite{Helenius:2017aqz}.

\begin{figure}
    \centering
    \includegraphics[width=0.48\textwidth]{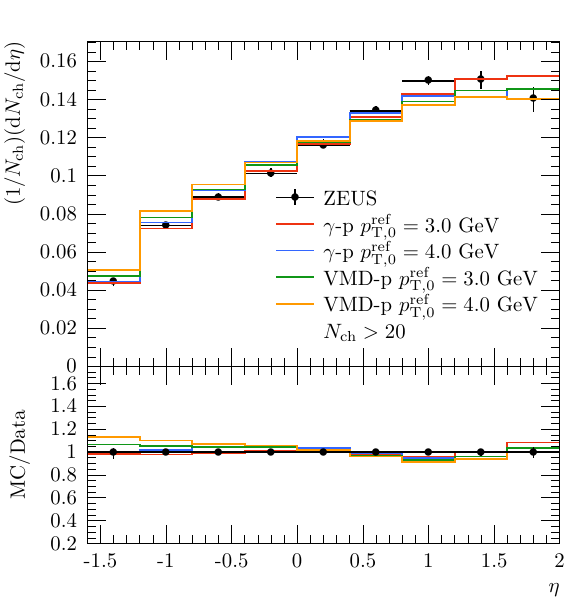}
    \includegraphics[width=0.48\textwidth]{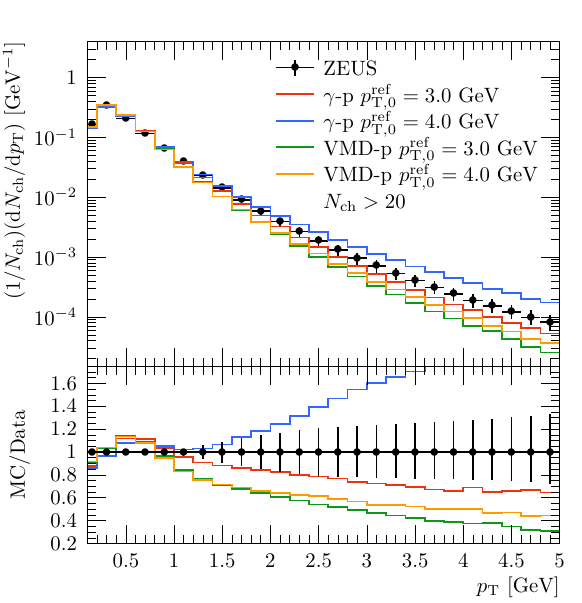}
    \includegraphics[width=0.48\textwidth]{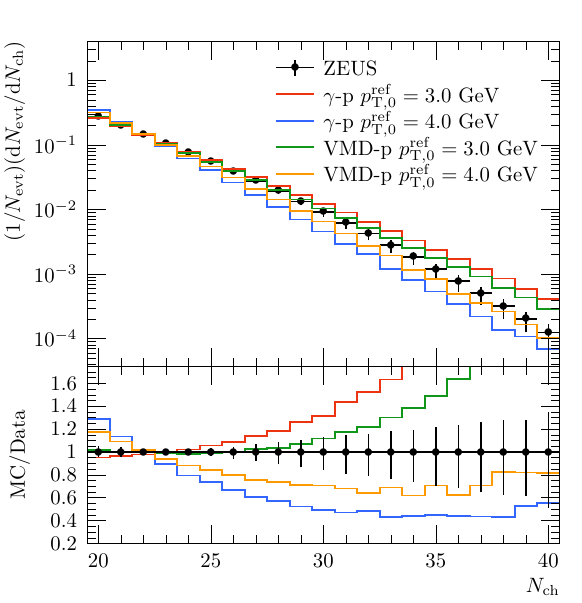}
    \includegraphics[width=0.48\textwidth]{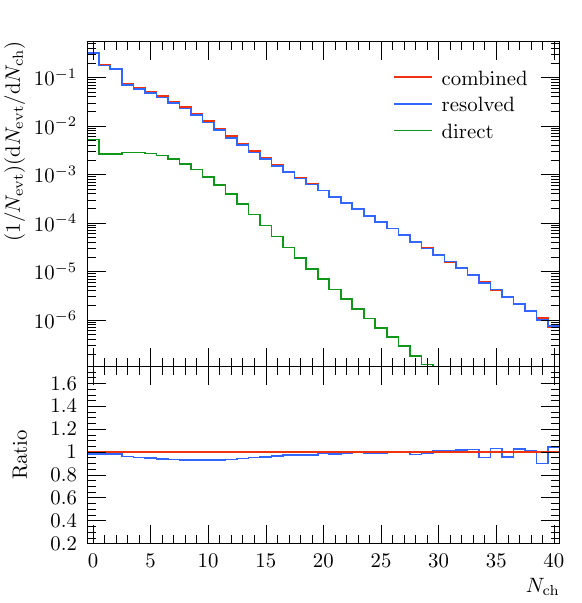}
    \caption{Pseudoradipity (upper left), $p_{\mathrm{T}}$ spectra (upper right) and multiplicity distribution (lower left) from full photoproduction and VMD model with different $p_{\mathrm{T,0}}^{\mathrm{ref}}$ compared to ZEUS data taken from electron-proton collisions at $\sqrt{s} = 318~\mathrm{GeV}$ \cite{ZEUS:2021qzg}, and multiplicity distribution with contributions from direct- and resolved-photon originated processes (lower right).} 
    \label{fig:ZEUS}
\end{figure}

In \figref{fig:ZEUS} we compare our \Pythia simulations to the ZEUS measurement for charged-particle multiplicity, $N_{\mathrm{ch}}$, and pseudorapidity, $\eta$, distribution and the $p_{\mathrm{T}}$ spectrum. The simulations include results with full photoproduction and with only the resolved part modelled with the VMD. In both cases we have varied $p_{\mathrm{T,0}}^{\mathrm{ref}}$ to account for the underlying uncertainty in the modelling of MPIs. In case of $N_{\mathrm{ch}}$ we notice that both full photoproduction and VMD calculation enclose the data within the $p_{\mathrm{T,0}}^{\mathrm{ref}}$ variation and that impact of this variation is somewhat smaller in case of VMD setup with a slight preference on $p_{\mathrm{T,0}}^{\mathrm{ref}} = 3~\text{GeV}$. In case of rapidity distribution the variations with different contributions are found small and well in line with the data. Only in case of the $p_{\mathrm{T}}$ spectrum do we find significant differences between full photoproduction and VMD only, where the former is again enclosing the data within $p_{\mathrm{T,0}}^{\mathrm{ref}}$ variations whereas the latter seem to fall below the data at high values of $p_{\mathrm{T}}$. This likely follows from differences in PDFs for vector mesons and resolved photons, as the latter tend to have more large-$x$ partons. Additionally we show also the break down of events from the different contributions for the multiplicity distributions with the default photoproduction setup in \figref{fig:ZEUS}. We find that when event selection does not require any high-$p_{\mathrm{T}}$ particles, the direct contribution is at most around 10 \% and negligible at high multiplicities. We conclude that the VMD-only setup is sufficient to describe the event structure for high-multiplicity events we focus on this study and that it provides a reasonable baseline to extend this to collisions with a heavy-ion target.

\subsection{Ultra-peripheral photo-nuclear collisions at the LHC}
\label{sec:UPCcomp}

The main application of our hadron-ion modelling is to enable simulations of photon-ion collisions applying the VMD model within \Pythia. Here we focus on high-multiplicity UPC events recently analysed by the ATLAS collaboration \cite{ATLAS:2021jhn}. The caveat of the data provided by the experimental analysis is that the shown distributions are not corrected for the finite acceptance and only the Fourier coefficients fitted to two-particle correlations are publicly available. This prevents direct comparisons, but in order to draw at least qualitative conclusions based on the data, we have adjusted the calculated MC multiplicity, $N_{\mathrm{ch}}$, to the experimentally reconstructed charged-particle multiplicity $N^{\mathrm{rec}}_{\mathrm{ch}}$ by scaling the former with a factor 0.8. This acceptance correction was estimated based on the \Pythia simulations for $\gamma$-p presented in ATLAS publication including the limited acceptance and matching this with simulations including all particles with similar event selection and \Pythia configuration. Similar efficiencies have been reported earlier by ATLAS in other collisions systems \cite{ATLAS:2016yzd} though instead of a constant factor some dependence on $\eta$ was observed.

\begin{figure}
    \centering
    \includegraphics[width=0.48\textwidth]{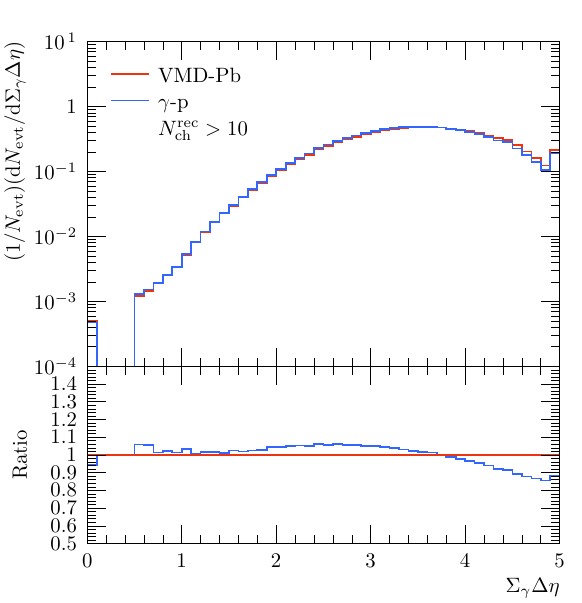}
    \includegraphics[width=0.48\textwidth]{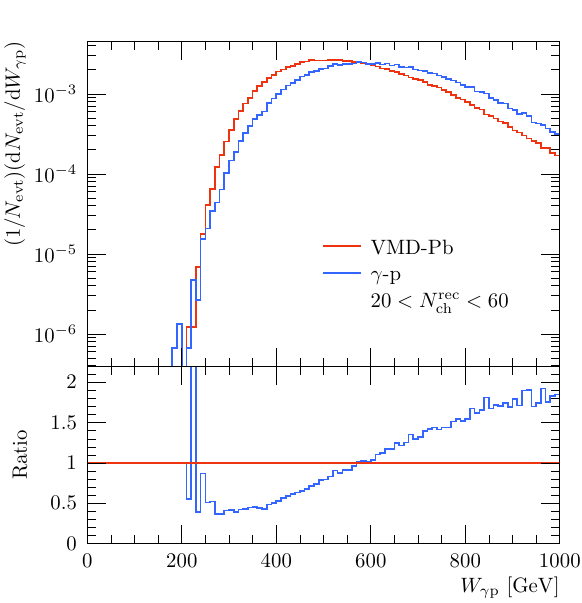}
    \caption{Rapidity gap measure applied by ATLAS \cite{ATLAS:2021jhn} (left) and invariant mass of the vector meson-nucleon system (right) from VMD-Pb configuration and photoproduction with proton target from \Pythia.}
    \label{fig:UPC-default}
\end{figure}

In addition to the efficiency correction, we perform a similar event selection criteria as quoted by the experimental study including a cut on sum-of-rapidity-gap measure $\Sigma_{\gamma}\Delta \eta$ of 2.5. This is calculated by summing over all rapidity gaps larger than $0.5$ in the range $0 \leq \eta \leq 4.9$ in the photon-going direction. The resulting $\Sigma_{\gamma}\Delta \eta$ distributions with the default VMD-Pb and photoproduction setup with proton target are shown in \figref{fig:UPC-default} together with the invariant mass distribution of the $\gamma$-nucleon system, $W_{\gamma \mathrm{p}}$, when a multiplicity cut of $20 < N_{\mathrm{ch}}^{\mathrm{rec}} < 60$ is applied. In case of the rapidity-gap measure we notice that the results with nuclear and nucleon targets are close to each other which can be expected since the experimental data was well in line with the simulations with proton target in the $\Sigma_{\gamma}\Delta \eta > 2.5$ region where photon-originated processes dominate. Similarly there are no big differences in the distribution of the photon-nucleon centre-of-mass energy $W_{\gamma \mathrm{p}}$ when applying cuts for the event multiplicity. In both cases the minimum centre-of-mass energy to produce events with $N^{\mathrm{rec}}_{\mathrm{ch}} > 20$ is found to be around $200~\text{GeV}$.
The average $W_{\gamma \mathrm{p}}$ distribution is slightly lower than with the VMD-Pb setup compared to simulation with proton target only which follows from the increased multiplicity for events with nuclear target.
There is also a correlation between the event multiplicity and centre-of-mass energy as the average $W_{\gamma \mathrm{p}}$ is around $470~\text{GeV}$ for the reconstructed multiplicities at $15-20$ and at $570~\text{GeV}$ for $20 < N^{\mathrm{rec}}_{\mathrm{ch}} < 60$.

\begin{figure}
    \centering
    \includegraphics[width=0.48\textwidth]{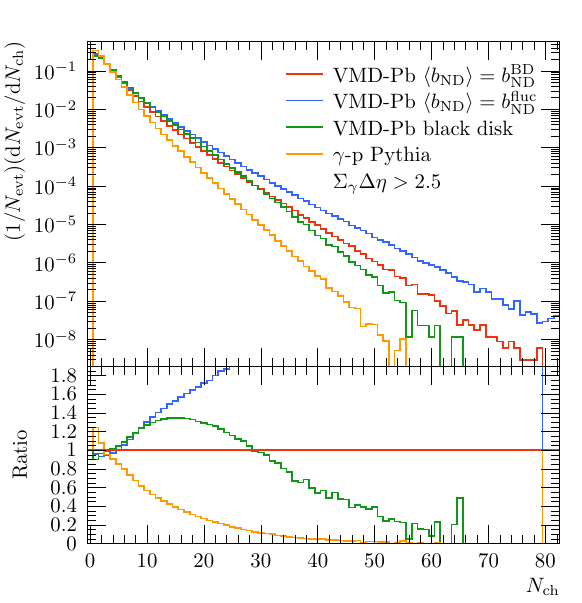}
    \includegraphics[width=0.48\textwidth]{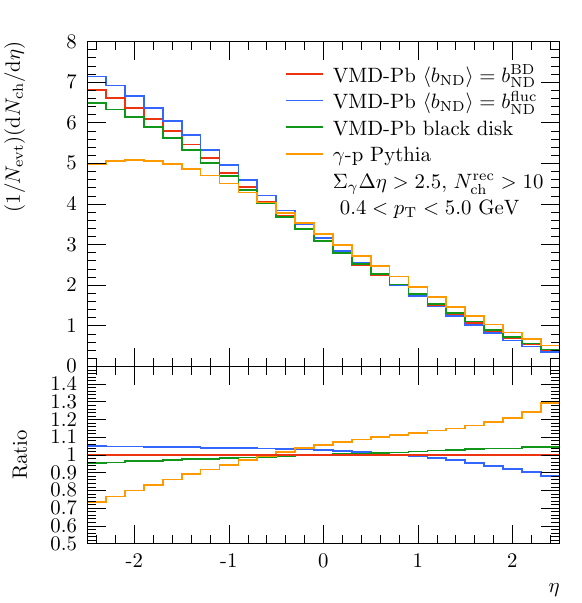}
    \caption{\Pythia results for per-event multiplicity (left) and rapidity distribution (right). For the VMD-Pb configuration results with different assumptions for \Angantyr subcollision model are compared to photoproduction with proton target.}
    \label{fig:UPCcollmod} 
\end{figure}

In \figref{fig:UPCcollmod} we include results with different subcollisions models included within the \Angantyr setup. Two of these include the cross section fluctuations with an average impact-parameter for non-diffractive events taken from the black-disk approximation ($b_{\mathrm{ND}}^{\mathrm{BD}}$) or from the fluctuating default model  ($b_{\mathrm{ND}}^{\mathrm{fluc}}$). In the black-disk model the nucleon radius and cross section fractions are fixed. 
To have a baseline to which the effects from a nuclear target can be compared, we also include results using the default photoproduction setup for $\gamma$-p that corresponds to the \Pythia results in the ATLAS publication \cite{ATLAS:2021jhn}. We present results for per-event yields as a function of $N_{\mathrm{ch}}$, and $\eta$. With all available nuclear collision models we notice that the multiplicities are increased compared to the simulation with just a proton target which is explained by subsequent interactions between the projectile and target nucleus. 

For the two models including cross section fluctuations we find similar shape in the multiplicity distribution, but somewhat lower tail of high-multiplicity events in case $\langle b_{\mathrm{ND}} \rangle$ calculated from the black-disk approximation. Keeping in mind that the multiplicities shown by ATLAS were not corrected for the limited efficiency, the simulated multiplicity distributions including fluctuations seem to enclose the measured distribution. However, the fixed-radii black-disk model high-multiplicity tail seems to fall short compared to data.

The observations are very similar for the pseudorapidity distributions where the fluctuating subcollision models predict a larger increase of particle production at negative rapidities as the fixed-radii model and the results with only a proton target. Again the data would seem to prefer models including cross-section fluctuations but, as above, the observations are only indicative due to missing unfolded data.

\begin{figure}
    \centering
    \includegraphics[width=0.48\textwidth]{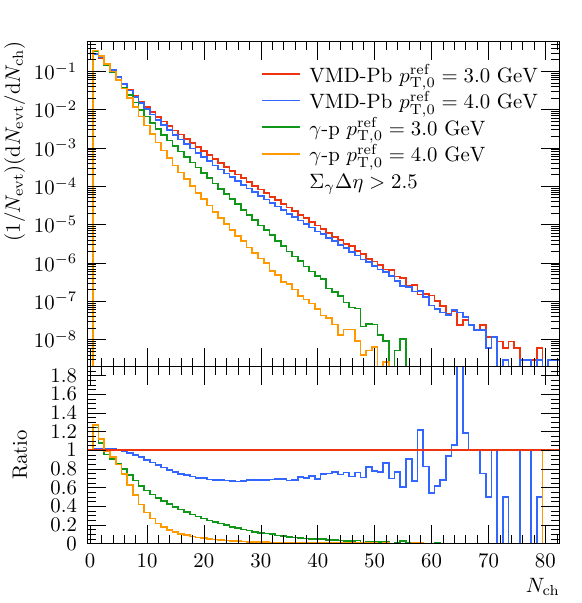}
    \includegraphics[width=0.48\textwidth]{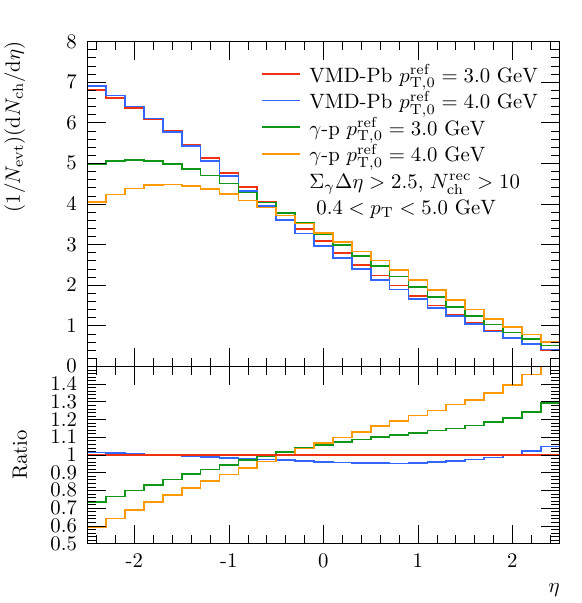}
    \caption{\Pythia results for per-event multiplicity (left) and rapidity distribution (right). \Pythia results include simulations with two different values for $p^{\mathrm{ref}}_{\mathrm{T},0}$ in case of VMD-Pb with the \Angantyr and photoproduction with proton target.}
    \label{fig:UPCpT0} 
\end{figure}

In \figref{fig:UPCpT0} we compare the simulations with varying $\pTRef$ using similar values for the parameter as in case of HERA data in \figref{fig:ZEUS} using the fluctuating subcollision model with $\langle b_{\mathrm{ND}} \rangle$ from black-disk approximation and comparing this with the results from applying photoproduction in $\gamma$-p. For the multiplicity distribution we find some dependence on this MPI parameter, but interestingly the effect is significantly reduced compared to the effect seen with a proton target. The reason for this is that now the multiplicity of an event is not only driven by the number of MPIs in a single collision between the projectile and a nucleon, but also by the number of collisions between the projectile and target nucleons. The latter seems to be the dominant mechanism at high multiplicities but some effects from the $\pTRef$ variation is seen around $10 < N^{\mathrm{rec}}_{\mathrm{ch}} < 50$. Similarly, only small effects from $\pTRef$ variation are found for the $\eta$ distribution considered in \figref{fig:UPCpT0}. Thus we conclude that while the probability of MPIs in case of real photons has some uncertainty, the results with a heavy-ion target are fairly robust against this.

\subsection{Collectivity in $\gamma$-Pb}
\label{sec:vnn}

\begin{figure}
    \centering
    \includegraphics[width=\textwidth]{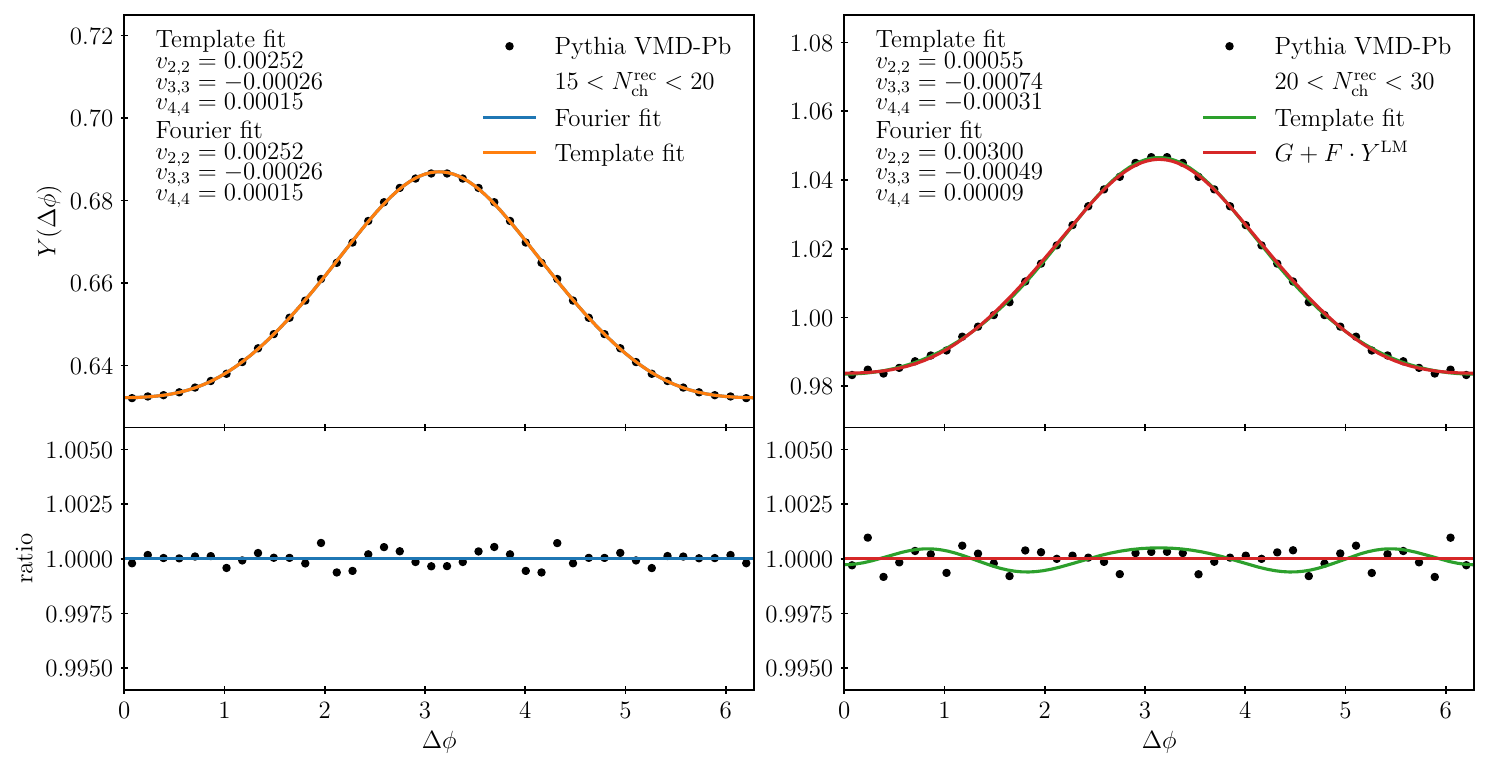}
    \caption{A Template fit to two-particle yields for low- (left) and high-multiplicity events (right) following the fit procedure in Ref.~\cite{ATLAS:2021jhn}. The multiplicity cuts have been adjusted to account for limited detector acceptance, see text for details.}
    \label{fig:dPhiFit} 
\end{figure}

Having a framework to simulate complete UPC events we can apply this also to study many-particle correlations. A particularly interesting observation in the ATLAS analysis \cite{ATLAS:2021jhn} was a finite value for the Fourier coefficients in high-multiplicity UPC events from the template fitting where the correlations in low-multiplicity events are subtracted from the high-multiplicity ones. In the UPCs, there is a clear correlation between the collision energy $W$ and the multiplicity, so it is not trivial whether the template fit would result as zero $v_n$ even if there are no collectivity from explicit final-state interactions. To verify this in detail, we have performed a similar template fit with the simulated events, using again the fluctuating subcollision model with $\langle b_{\mathrm{ND}} \rangle = b_{\mathrm{ND}}^{\mathrm{BD}}$. The calculated two-particle yield as a function of the azimuth angle $\phi$ and pseudorapidity separation between the two particles $|\Delta \eta|$, $Y(\Delta \phi, |\Delta \eta|)$, is integrated over $2.0 < |\Delta \eta| < 5.0$ to obtain the one-dimensional two-particle yield $Y(\Delta \phi)$. For the low-multiplicity (LM) events, a direct Fourier fit is performed by accounting terms up to order 4 in the Fourier series of the form
\begin{equation}
Y^{\mathrm{LM}}(\Delta \phi) = c_0 + 2 \sum_{n=1}^4 c_n \cos(n \cdot \Delta \phi).
\label{eq:Ffit}
\end{equation}
For high-multiplicity (HM) events a parametrization is used,
\begin{equation}
Y^{\mathrm{HM}}(\Delta \phi) = F \cdot Y^{\mathrm{LM}}(\Delta \phi) + G \left[ 1 +  2 \sum_{n=2}^4 v_{n,n} \cos(n \cdot \Delta \phi)\right],
\label{eq:Tfit}
\end{equation}
where now $F$ and $G$ are free parameters, but are connected by the requirement of having the same value for the integral of both sides in equation (\ref{eq:Tfit}). In total there are then 9 free parameters that are fitted simultaneously. In addition, we perform a direct Fourier fit to the HM sample using the truncated series in equation (\ref{eq:Ffit}) to extract the Fourier coefficients without the template fitting and a separate 5-parameter Fourier fit to LM sample to check whether the simultaneous template fit modifies the values of the Fourier coefficients. An example of this procedure is shown in \figref{fig:dPhiFit} where the LM sample has $15 < N^{\mathrm{rec}}_{\mathrm{ch}} < 20$, the HM sample has $20 < N^{\mathrm{rec}}_{\mathrm{ch}} < 30$ and the $p_{\mathrm{T}}$ of the both particles is between $0.4~\text{GeV}$ and $2.0~\text{GeV}$. In case of a direct Fourier fit we identify $c_n = v_{n,n}$. For the HM fit we plot also part $G + F \cdot Y^{\mathrm{LM}}(\Delta \phi)$ which provides the scaled LM reference to which the additional HM modulation can be compared to. The simulated results have been obtained with $\pTRef$ using the default subcollision model for the VMD-Pb simulations. We notice that for the LM sample both the template fit and the direct Fourier fit yields the same values for the parameters confirming that the LM events are equally well described by both fits. In case of HM events we find that the scaled modulation from LM events is sufficient to describe the simulated two-particle yield as the values of Fourier coefficients $v_{n,n}$ obtained from the Template fit are consistent with zero when accounting for the fit uncertainties related to finite statistics in the simulated events. This is also reflected by the ratio of the simulated events and the $G + F \cdot Y^{\mathrm{LM}}(\Delta \phi)$ where no further significant modulation remains. We find also that the Fourier coefficients from a direct Fourier fit are very similar in both event samples further reflecting similarity of the azimuthal modulation in LM and HM samples.

\begin{figure}
    \centering
    \includegraphics[width=0.48\textwidth]{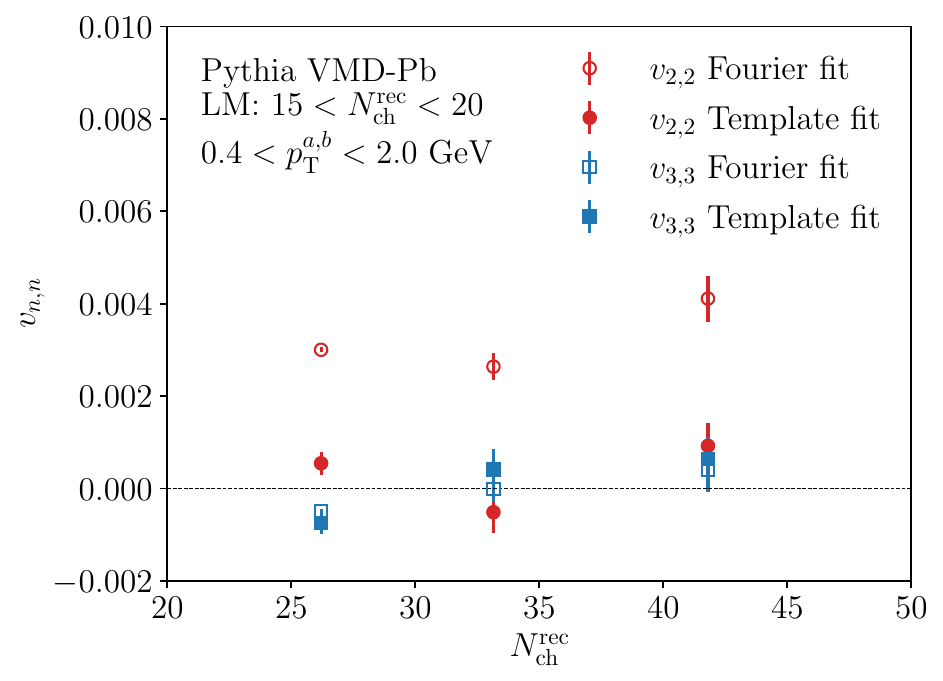}
    \includegraphics[width=0.48\textwidth]{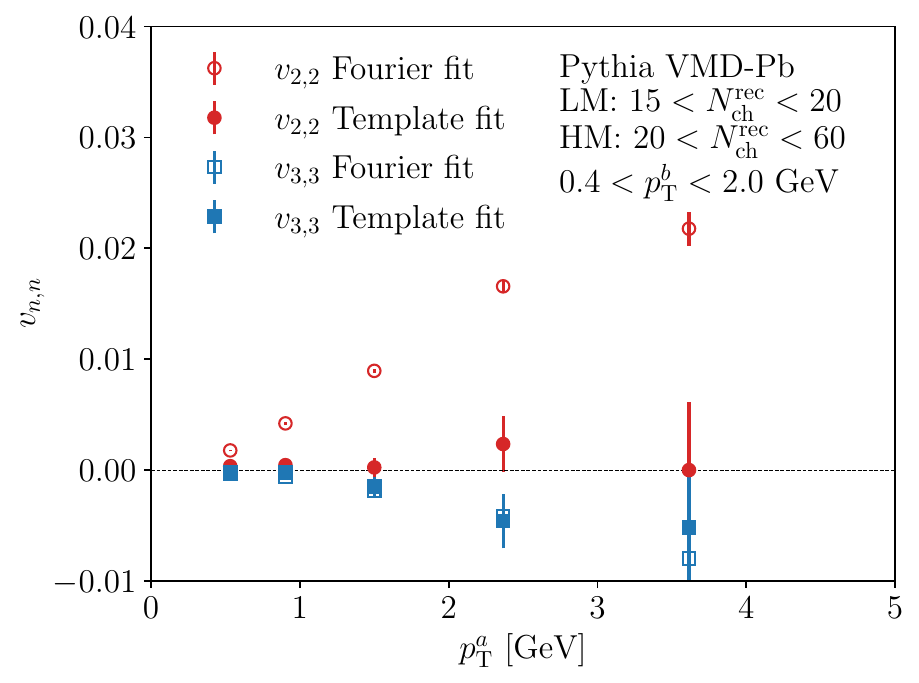}
    \caption{The $v_{n,n}$ values from the fit to \Pythia events at different multiplicities (left) and different $p^a_{\mathrm{T}}$ bins (right) corresponding to ATLAS study in Ref.~\cite{ATLAS:2021jhn}. Error bars quantify the statistical uncertainty in the fit.}
    \label{fig:vnn-err} 
\end{figure}

As in the ALTAS UPC analysis \cite{ATLAS:2021jhn}, we have repeated the fitting procedure for different selections of high-multiplicity events and by varying the transverse momentum of the particle $a$. The similar event selection criteria as in the ATLAS analysis have been applied accounting also for the limited efficiency for $N^{\mathrm{ch}}_{\mathrm{rec}}$ discussed above. The multiplicity and $p^{a}_{\mathrm{T}}$ dependencies are shown in \figref{fig:vnn-err}. For the direct Fourier fits (open markers) we notice that the trends are very similar with the experimental results. The fitted $v_{2,2}$ are of the same order and roughly constant as a function of $N^{\mathrm{ch}}_{\mathrm{rec}}$ and increase linearly as a function of $p^{a}_{\mathrm{T}}$. In case of $v_{3,3}$ the values are consistent with zero apart from the pairs with the highest $p^{a}_{\mathrm{T}}$ values where the negative $v_{3,3}$ are found, similar to the experimental data. In case of the template fit (filled markers) the clear difference to the measurement is that in all cases the resulting $v_{n,n}$ values are consistent with zero whereas in the data finite and positive values for $v_{2,2}$ and $v_{3,3}$ were found from the template fit. This further confirms the observation from the result in \figref{fig:dPhiFit} showing no room for additional modulation in HM events wrt. LM sample in our simulations. We conclude that our \Pythia VMD-Pb simulations without any explicit final-state interactions is not compatible with the ATLAS data where finite positive values for $v_{n,n}$ were found after the template fit.

\section{Summary and outlook}
\label{sec:summary}

In this article, we have presented a framework for hadron-ion collisions in \Pythia/\Angantyr where the allowed projectile hadrons include pions, vector mesons and quarkonia. The full list of available hadrons match the previous study related to hadron-nucleon collisions \cite{Sjostrand:2021dal}. The application considered in this study is modelling of photon-ion interactions in ultra-peripheral heavy-ion collisions using the vector-dominance model for the quasi-real photon. This framework can be also applied to study air showers created by high-energy cosmic rays when hitting to the Earth's atmosphere. 

The implementation is built upon \Pythia's existing framework for hadron-nucleon interactions, using partial hadronic cross sections as input for the Glauber subcollision modelling, and invoking \Pythia to simulate hadron-nucleon interactions. The component that requires most new modelling is the hadronic fluctuations, which are calculated by a genetic algorithm fitting to partial cross sections. In existing \Angantyr pA/AA, the fluctuation parameters are the same for both beams, and we extended this to include a model with asymmetric fluctuations for the general $h$A case. However, we found that using the asymmetric model gives only a small difference, especially compared to existing model uncertainties such as in the cross sections and PDFs for uncommon hadrons. Considering that it takes significantly longer to run the genetic algorithm with additional parameters, we believe the symmetric model gives sufficient precision for now. In the future, a more accurate modelling of these fluctuations may be relevant e.g. for the tail shape of multiplicity distributions. 

With the current fluctuation model, some hadrons (e.g. $J/\psi$) would fluctuate to unreasonably large sizes, contradicting the expectation that it should have a narrower wavefunction. This cannot be regulated in the current model because the fit is only to the inclusive cross sections, and it would require making deeper changes to the approach in order to accurately model different hadronic fluctuations. As a consequence to these large fluctuations, the mean impact parameter $\langle b_\mathrm{ND} \rangle$ also becomes very large, which causes the MPI framework to treat all collisions as essentially head-on, leading to too high MPI activity. As a workaround, we presented the option to use the black disk value, $\langle b_\mathrm{ND} \rangle = \frac23 \sqrt{\sigabs/\pi}$, but it is clear that additional modelling in necessary to accurately represent fluctuations of generic hadrons.

This development also includes a technical feature for changing the energy on an event-by-event basis in \Angantyr. For varying energies, one of the initialization bottlenecks is the genetic algorithm for hadronic fluctuations. We reduce the initialization time by saving the resulting parameters to disk at certain energies, and interpolating between these values whenever the energy is changed. Our study indicates that the parameters change slowly enough that roughly one interpolation point per order of magnitude is sufficient. As we move to energies below $\sim 50$~GeV, a growing fraction of events will include non-perturbative effects such as resonance formation. These interactions are implemented in \Pythia but not yet in \Angantyr, and it is not clear how accurate our framework is below this energy.

In addition to switching energies, we now also allow switching the beam particles. This is relevant for example in VMD models where the species depends on which vector meson the photon fluctuates into, and in cosmic rays where the hadronic cascades includes a wide variety of species. When switching beams, the projectile is allowed to be any hadron or nucleus, while the target can be a proton, neutron or nucleus.

We compared the VMD model first in case of photon-proton collisions with the full photoproduction implementation in \Pythia and with the experimental data from ZEUS for high-multiplicity photoproduction events. For the applied event selection the two models were in a good agreement with each other and the experimental data. In case of the multiplicity distribution we noticed significant differences when varying the value $\pTRef$ that controls the probability for the MPIs. With both models the data was, however, enclosed with $\pTRef$ range in line with previous studies. In case of photon-lead UPCs we found that the multiplicity distributions were not very sensitive to the value of this parameter, but the increased multiplicity when applying a nuclear target was rather driven by the number of collisions between the resolved photon and different nucleons in the nucleus. We found some dependence on the applied fluctuating cross-section model in \Angantyr and noticed that the ATLAS data do seem to disfavour simulations when these fluctuations are completely turned off and a simplified black disk model is applied. We conclude that the presented framework for photon-ion collisions do have some uncertainties related to subcollision modelling but the generic features are in line with the limited data and future measurements with more direct comparisons would provide useful constraints.

Having a model that can produce complete final states in photon-ion collisions, we were able to test whether the observed collective behaviour in two-particle correlations could arise from simulations without including any explicit final-state interactions. We noticed that while the direct Fourier fits to the obtained two-particle correlations where in line with ATLAS data, the $v_{n,n}$ values from the template fit were consistent with zero. In case of the ATLAS data, they found finite values for these Fourier coefficients after subtracting correlations from the low-multiplicity events with the template fitting procedure. This suggests that further final-state interactions are required to match the experimental values for the coefficients quantifying collective behaviour. In case of \Pythia context, such final-state effects could arise from string interactions such as rope hadronization and string shoving that we plan to revisit in a future study.

\section*{Acknowledgments}

We thank Leif Lönnblad and Christian Bierlich for useful discussions related to the \Angantyr model and Sruthy Das for information related to ATLAS particle-reconstruction efficiency. We also thank Maximilian Reininghaus for sharing cross section data from the Sibyll, EPOS-LHC and QGSJet models. We acknowledge the financial support from the Research Council of Finland, Project No. 331545, and through the Centre of Excellence in Quark Matter. The reported work is associated with the European Research Council project ERC-2018-ADG-835105 YoctoLHC. We acknowledge grants of computer capacity from the Finnish Grid and Cloud Infrastructure (persistent identifier\\ \texttt{urn:nbn:fi:research-infras-2016072533}).

\bibliographystyle{unsrt}
\bibliography{Angantyr-VMD}

\begin{thebibliography}{10}

\bibitem{Bierlich:2022pfr}
Christian Bierlich et~al.
\newblock {A comprehensive guide to the physics and usage of PYTHIA 8.3}.
\newblock {\em SciPost Phys. Codeb.}, 3 2022.

\bibitem{Collins:1989gx}
John~C. Collins, Davison~E. Soper, and George~F. Sterman.
\newblock {Factorization of Hard Processes in QCD}.
\newblock {\em Adv. Ser. Direct. High Energy Phys.}, 5:1--91, 1989.

\bibitem{Kovarik:2019xvh}
Karol Kova\v{r}\'\i{}k, Pavel~M. Nadolsky, and Davison~E. Soper.
\newblock {Hadronic structure in high-energy collisions}.
\newblock {\em Rev. Mod. Phys.}, 92(4):045003, 2020.

\bibitem{Hoche:2014rga}
Stefan H\"oche.
\newblock {Introduction to parton-shower event generators}.
\newblock In {\em {Theoretical Advanced Study Institute in Elementary Particle
  Physics}: {Journeys Through the Precision Frontier: Amplitudes for
  Colliders}}, pages 235--295, 2015.

\bibitem{Helenius:2017aqz}
Ilkka Helenius.
\newblock {Photon-photon and photon-hadron processes in Pythia 8}.
\newblock {\em CERN Proc.}, 1:119, 2018.

\bibitem{Bierlich:2018xfw}
Christian Bierlich, G\"osta Gustafson, Leif L\"onnblad, and Harsh Shah.
\newblock {The Angantyr model for Heavy-Ion Collisions in PYTHIA8}.
\newblock {\em JHEP}, 10:134, 2018.

\bibitem{Sjostrand:2021dal}
Torbj\"orn Sj\"ostrand and Marius Utheim.
\newblock {Hadron interactions for arbitrary energies and species, with
  applications to cosmic rays}.
\newblock {\em Eur. Phys. J. C}, 82(1):21, 2022.

\bibitem{Accardi:2012qut}
A.~Accardi et~al.
\newblock {Electron Ion Collider: The Next QCD Frontier}: {Understanding the
  glue that binds us all}.
\newblock {\em Eur. Phys. J. A}, 52(9):268, 2016.

\bibitem{AbdulKhalek:2021gbh}
R.~Abdul~Khalek et~al.
\newblock {Science Requirements and Detector Concepts for the Electron-Ion
  Collider}: {EIC Yellow Report}.
\newblock {\em Nucl. Phys. A}, 1026:122447, 2022.

\bibitem{Bertulani:2005ru}
Carlos~A. Bertulani, Spencer~R. Klein, and Joakim Nystrand.
\newblock {Physics of ultra-peripheral nuclear collisions}.
\newblock {\em Ann. Rev. Nucl. Part. Sci.}, 55:271--310, 2005.

\bibitem{Klein:2020fmr}
Spencer Klein and Peter Steinberg.
\newblock {Photonuclear and Two-photon Interactions at High-Energy Nuclear
  Colliders}.
\newblock {\em Ann. Rev. Nucl. Part. Sci.}, 70:323--354, 2020.

\bibitem{Klasen:2002xb}
Michael Klasen.
\newblock {Theory of hard photoproduction}.
\newblock {\em Rev. Mod. Phys.}, 74:1221--1282, 2002.

\bibitem{Butterworth:2005aq}
J.~M. Butterworth and M.~Wing.
\newblock {High energy photoproduction}.
\newblock {\em Rept. Prog. Phys.}, 68:2773--2828, 2005.

\bibitem{ALICE:2013wjo}
E.~Abbas et~al.
\newblock {Charmonium and $e^+e^-$ pair photoproduction at mid-rapidity in
  ultra-peripheral Pb-Pb collisions at $\sqrt{s_{\rm NN}}$=2.76 TeV}.
\newblock {\em Eur. Phys. J. C}, 73(11):2617, 2013.

\bibitem{ALICE:2021gpt}
Shreyasi Acharya et~al.
\newblock {Coherent $J/\psi$ and $\psi'$ photoproduction at midrapidity in
  ultra-peripheral Pb-Pb collisions at $\sqrt{s_{\mathrm{NN}}}~=~5.02$ TeV}.
\newblock {\em Eur. Phys. J. C}, 81(8):712, 2021.

\bibitem{CMS:2016itn}
Vardan Khachatryan et~al.
\newblock {Coherent $J/\psi$ photoproduction in ultra-peripheral PbPb
  collisions at $\sqrt {s_{NN}} =$ 2.76 TeV with the CMS experiment}.
\newblock {\em Phys. Lett. B}, 772:489--511, 2017.

\bibitem{LHCb:2021bfl}
Roel Aaij et~al.
\newblock {Study of coherent $J/\psi$ production in lead-lead collisions at $
  \sqrt{{\mathrm{s}}_{\mathrm{NN}}} $ = 5 TeV}.
\newblock {\em JHEP}, 07:117, 2022.

\bibitem{ATLAS:2022cbd}
ATLAS collaboration (2022).
\newblock {Photo-nuclear jet production in ultra-peripheral Pb+Pb collisions at
  $\sqrt{s}_\text{NN} = 5.02$ TeV with the ATLAS detector}.
\newblock {\em Contribution to: Quark Matter}, 2022.

\bibitem{ATLAS:2021jhn}
Georges Aad et~al.
\newblock {Two-particle azimuthal correlations in photonuclear ultraperipheral
  Pb+Pb collisions at 5.02 TeV with ATLAS}.
\newblock {\em Phys. Rev. C}, 104(1):014903, 2021.

\bibitem{Heinz:2013th}
Ulrich Heinz and Raimond Snellings.
\newblock {Collective flow and viscosity in relativistic heavy-ion collisions}.
\newblock {\em Ann. Rev. Nucl. Part. Sci.}, 63:123--151, 2013.

\bibitem{CMS:2015yux}
Vardan Khachatryan et~al.
\newblock {Evidence for Collective Multiparticle Correlations in p-Pb
  Collisions}.
\newblock {\em Phys. Rev. Lett.}, 115(1):012301, 2015.

\bibitem{ATLAS:2016yzd}
Morad Aaboud et~al.
\newblock {Measurements of long-range azimuthal anisotropies and associated
  Fourier coefficients for $pp$ collisions at $\sqrt{s}=5.02$ and $13$ TeV and
  $p$+Pb collisions at $\sqrt{s_{\mathrm{NN}}}=5.02$ TeV with the ATLAS
  detector}.
\newblock {\em Phys. Rev. C}, 96(2):024908, 2017.

\bibitem{ATLAS:2015hzw}
Georges Aad et~al.
\newblock {Observation of Long-Range Elliptic Azimuthal Anisotropies in
  $\sqrt{s}=$13 and 2.76 TeV $pp$ Collisions with the ATLAS Detector}.
\newblock {\em Phys. Rev. Lett.}, 116(17):172301, 2016.

\bibitem{ATLAS:2019wzn}
Morad Aaboud et~al.
\newblock {Measurement of long-range two-particle azimuthal correlations in
  $Z$-boson tagged $pp$ collisions at $\sqrt{s}{=}8$ and 13 TeV}.
\newblock {\em Eur. Phys. J. C}, 80(1):64, 2020.

\bibitem{Gluck:1991jc}
M.~Gluck, E.~Reya, and A.~Vogt.
\newblock {Photonic parton distributions}.
\newblock {\em Phys. Rev. D}, 46:1973--1979, 1992.

\bibitem{Sherpa:2019gpd}
Enrico Bothmann et~al.
\newblock {Event Generation with Sherpa 2.2}.
\newblock {\em SciPost Phys.}, 7(3):034, 2019.

\bibitem{Bellm:2015jjp}
Johannes Bellm et~al.
\newblock {Herwig 7.0/Herwig++ 3.0 release note}.
\newblock {\em Eur. Phys. J. C}, 76(4):196, 2016.

\bibitem{Hoeche:2023gme}
Stefan Hoeche, Frank Krauss, and Peter Meinzinger.
\newblock {Resolved photons in Sherpa}.
\newblock {\em Eur. Phys. J. C}, 84(2):178, 2024.

\bibitem{Helenius:2024rth}
Ilkka Helenius, Peter Meinzinger, Simon Pl\"atzer, and Peter Richardson.
\newblock {Photoproduction in general-purpose event generators}.
\newblock \texttt{arXiv:2406.08026 [hep-ph]} (2024).

\bibitem{ZEUS:2021qzg}
I.~Abt et~al.
\newblock {Azimuthal correlations in photoproduction and deep inelastic $ep$
  scattering at HERA}.
\newblock {\em JHEP}, 12:102, 2021.

\bibitem{CMS:2022doq}
Armen Tumasyan et~al.
\newblock {Two-particle azimuthal correlations in \ensuremath{\gamma}p
  interactions using pPb collisions at sNN=8.16TeV}.
\newblock {\em Phys. Lett. B}, 844:137905, 2023.

\bibitem{ParticleDataGroup:2022pth}
R.~L. Workman et~al.
\newblock {Review of Particle Physics}.
\newblock {\em PTEP}, 2022:083C01, 2022.

\bibitem{Engel:2018akg}
Ralph Engel, Dieter Heck, Tim Huege, Tanguy Pierog, Maximilian Reininghaus,
  Felix Riehn, Ralf Ulrich, Michael Unger, and Darko Veberič.
\newblock {Towards a Next Generation of {CORSIKA}: A Framework for the
  Simulation of Particle Cascades in Astroparticle Physics}.
\newblock {\em Comput. Softw. Big Sci.}, 3(1):2, 2019.

\bibitem{Engel:2019dsg}
Felix Riehn, Ralph Engel, Anatoli Fedynitch, Thomas~K. Gaisser, and Todor
  Stanev.
\newblock {Hadronic interaction model Sibyll 2.3d and extensive air showers}.
\newblock {\em Phys. Rev. D}, 102(6):063002, 2020.

\bibitem{Pierog:2013ria}
T.~Pierog, Iu. Karpenko, J.~M. Katzy, E.~Yatsenko, and K.~Werner.
\newblock {EPOS LHC: Test of collective hadronization with data measured at the
  CERN Large Hadron Collider}.
\newblock {\em Phys. Rev. C}, 92(3):034906, 2015.

\bibitem{Ostapchenko:2010vb}
Sergey Ostapchenko.
\newblock {Monte Carlo treatment of hadronic interactions in enhanced Pomeron
  scheme: I. QGSJET-II model}.
\newblock {\em Phys. Rev. D}, 83:014018, 2011.

\bibitem{Reininghaus:2023ctx}
Maximilian Reininghaus, Torbj\"orn Sj\"ostrand, and Marius Utheim.
\newblock {Pythia 8 as hadronic interaction model in air shower simulations}.
\newblock {\em EPJ Web Conf.}, 283:05010, 2023.

\bibitem{Bierlich:2021poz}
Christian Bierlich, Torbj\"orn Sj\"ostrand, and Marius Utheim.
\newblock {Hadronic rescattering in pA and AA collisions}.
\newblock {\em Eur. Phys. J. A}, 57(7):227, 2021.

\bibitem{Donnachie:1992ny}
A.~Donnachie and P.~V. Landshoff.
\newblock {Total cross-sections}.
\newblock {\em Phys. Lett. B}, 296:227--232, 1992.

\bibitem{Lipkin:1973nt}
Harry~J. Lipkin.
\newblock {Quarks for pedestrians}.
\newblock {\em Phys. Rept.}, 8:173--268, 1973.

\bibitem{Levin:1965mi}
E.~M. Levin and L.~L. Frankfurt.
\newblock {The Quark hypothesis and relations between cross-sections at
  high-energies}.
\newblock {\em JETP Lett.}, 2:65--70, 1965.

\bibitem{Schuler:1993wr}
Gerhard~A. Schuler and Torbjorn Sjostrand.
\newblock {Hadronic diffractive cross-sections and the rise of the total
  cross-section}.
\newblock {\em Phys. Rev. D}, 49:2257--2267, 1994.

\bibitem{Schuler:1996en}
Gerhard~A. Schuler and Torbjorn Sjostrand.
\newblock {A Scenario for high-energy gamma gamma interactions}.
\newblock {\em Z. Phys. C}, 73:677--688, 1997.

\bibitem{Gluck:1991ey}
M.~Gl\"uck, E.~Reya, and A.~Vogt.
\newblock {Pionic parton distributions}.
\newblock {\em Z. Phys. C}, 53:651--656, 1992.

\bibitem{Gluck:1999xe}
M.~Gl\"uck, E.~Reya, and I.~Schienbein.
\newblock {Pionic parton distributions revisited}.
\newblock {\em Eur. Phys. J. C}, 10:313--317, 1999.

\bibitem{Glauber:1955qq}
R.~J. Glauber.
\newblock {Cross-sections in deuterium at high-energies}.
\newblock {\em Phys. Rev.}, 100:242--248, 1955.

\bibitem{Rybczynski:2013yba}
Maciej Rybczynski, Grzegorz Stefanek, Wojciech Broniowski, and Piotr Bozek.
\newblock {GLISSANDO 2 : GLauber Initial-State Simulation AND
  mOre\textellipsis{}, ver. 2}.
\newblock {\em Comput. Phys. Commun.}, 185:1759--1772, 2014.

\bibitem{Heiselberg:1991is}
H.~Heiselberg, G.~Baym, B.~Blaettel, L.~L. Frankfurt, and M.~Strikman.
\newblock {Color transparency, color opacity, and fluctuations in nuclear
  collisions}.
\newblock {\em Phys. Rev. Lett.}, 67:2946--2949, 1991.

\bibitem{Blaettel:1993ah}
B.~Blaettel, G.~Baym, L.~L. Frankfurt, H.~Heiselberg, and M.~Strikman.
\newblock {Hadronic cross-section fluctuations}.
\newblock {\em Phys. Rev. D}, 47:2761--2772, 1993.

\bibitem{Alvioli:2013vk}
M.~Alvioli and M.~Strikman.
\newblock {Color fluctuation effects in proton-nucleus collisions}.
\newblock {\em Phys. Lett. B}, 722:347--354, 2013.

\bibitem{Alvioli:2014sba}
M.~Alvioli, L.~Frankfurt, V.~Guzey, and M.~Strikman.
\newblock {Revealing \textquotedblleft{}flickering\textquotedblright{} of the
  interaction strength in pA collisions at the CERN LHC}.
\newblock {\em Phys. Rev. C}, 90:034914, 2014.

\bibitem{Alvioli:2014eda}
Massimiliano Alvioli, Brian~A. Cole, Leonid Frankfurt, D.~V. Perepelitsa, and
  Mark Strikman.
\newblock {Evidence for $x$-dependent proton color fluctuations in pA
  collisions at the CERN Large Hadron Collider}.
\newblock {\em Phys. Rev. C}, 93(1):011902, 2016.

\bibitem{Sjostrand:2004pf}
T.~Sjostrand and Peter~Z. Skands.
\newblock {Multiple interactions and the structure of beam remnants}.
\newblock {\em JHEP}, 03:053, 2004.

\bibitem{Bierlich:2016vgw}
Christian Bierlich, G\"osta Gustafson, and Leif L\"onnblad.
\newblock {A shoving model for collectivity in hadronic collisions}.
\newblock \texttt{arXiv:1612.05132 [hep-ph]} (2016).

\bibitem{Bierlich:2017vhg}
Christian Bierlich, G\"osta Gustafson, and Leif L\"onnblad.
\newblock {Collectivity without plasma in hadronic collisions}.
\newblock {\em Phys. Lett. B}, 779:58--63, 2018.

\bibitem{Bierlich:2020naj}
Christian Bierlich, Smita Chakraborty, G\"osta Gustafson, and Leif L\"onnblad.
\newblock {Setting the string shoving picture in a new frame}.
\newblock {\em JHEP}, 03:270, 2021.

\bibitem{Helenius:2019gbd}
Ilkka Helenius and Christine~O. Rasmussen.
\newblock {Hard diffraction in photoproduction with Pythia 8}.
\newblock {\em Eur. Phys. J. C}, 79(5):413, 2019.

\bibitem{Schuler:1992dt}
Gerhard~A. Schuler and Torbjorn Sjostrand.
\newblock {The hadronic properties of the photon in gamma p interactions}.
\newblock {\em Phys. Lett. B}, 300:169--174, 1993.

\bibitem{Bauer:1977iq}
T.~H. Bauer, R.~D. Spital, D.~R. Yennie, and F.~M. Pipkin.
\newblock {The Hadronic Properties of the Photon in High-Energy Interactions}.
\newblock {\em Rev. Mod. Phys.}, 50:261, 1978.
\newblock [Erratum: Rev.Mod.Phys. 51, 407 (1979)].

\bibitem{Schuler:1994ft}
Gerhard~A. Schuler and Torbjorn Sjostrand.
\newblock {Gamma gamma and gamma P events at high-energies}.
\newblock In {\em {Two-Photon Physics from DAPHNE to LEP200 and Beyond}}, 3
  1994.

\bibitem{Budnev:1975poe}
V.~M. Budnev, I.~F. Ginzburg, G.~V. Meledin, and V.~G. Serbo.
\newblock {The Two photon particle production mechanism. Physical problems.
  Applications. Equivalent photon approximation}.
\newblock {\em Phys. Rept.}, 15:181--281, 1975.

\bibitem{vonWeizsacker:1934nji}
C.~F. von Weizsacker.
\newblock {Radiation emitted in collisions of very fast electrons}.
\newblock {\em Z. Phys.}, 88:612--625, 1934.

\bibitem{Williams:1934ad}
E.~J. Williams.
\newblock {Nature of the high-energy particles of penetrating radiation and
  status of ionization and radiation formulae}.
\newblock {\em Phys. Rev.}, 45:729--730, 1934.

\bibitem{Skands:2014pea}
Peter Skands, Stefano Carrazza, and Juan Rojo.
\newblock {Tuning PYTHIA 8.1: the Monash 2013 Tune}.
\newblock {\em Eur. Phys. J. C}, 74(8):3024, 2014.

\end{thebibliography}

\end{document}